\documentclass[twoside,11pt]{article}
\usepackage{epsfig,rotating,cite}

\newcommand{\pp}       {${\rm p\bar{p}}$}
\font\aa=cmr7
\aa 

\begin{document}

\begin{titlepage}
\begin{flushleft}
{\tt DESY 97-075}\hfill {\tt ISSN 0418-9833} \\
\end{flushleft}

\begin{center}
\vspace*{1.5cm}
\begin{huge}
{\bf Bose-Einstein Correlations \linebreak[4] in \linebreak[4] 
Deep Inelastic \mbox{\boldmath{$ep$}} Scattering at HERA\\[3.cm]} 
\vspace*{0.5cm}
\end{huge}

\begin{Large}
H1 Collaboration \\
\end{Large}
\end{center}
\vspace*{3cm}
\nopagebreak

\begin{abstract} 
\small
\noindent
Two-particle correlations in invariant mass are studied separately for
like-sign and unlike-sign charged particles produced in deep inelastic
positron-proton scattering in a new kinematical domain. 
The data were taken with the H1 detector at the HERA storage ring in
1994, in which 27.5 GeV positrons collided with 820 GeV protons at
a centre of mass energy $\sqrt{s}=300$ GeV.

The observed enhancement of the like-sign correlations at low invariant masses 
is related to the dimensions of the hadronic source.  
The data are compared to different QCD models
where the hadronization is performed with the string-fragmentation
model. Results are presented for the first time
separately for diffractive and non-diffractive scattering, 
in domains of four-momentum transfer, Bjorken-$x$, and hadronic center 
of mass energy, and in intervals of charged particle multiplicity.  
The observed source radii do not differ strongly
from those measured in lower energy 
lepton-nucleon inelastic scattering, and $e^+e^-$ annihilation.

\normalsize
\end{abstract}
\vfill
\end{titlepage}
 
\newpage
\begin{flushleft}
\setlength{\parindent}{0mm}

 C.~Adloff$^{35}$,                
 S.~Aid$^{13}$,                   
 M.~Anderson$^{23}$,              
 V.~Andreev$^{26}$,               
 B.~Andrieu$^{29}$,               
 V.~Arkadov$^{36}$,               
 C.~Arndt$^{11}$,                 
 I.~Ayyaz$^{30}$,                 
 A.~Babaev$^{25}$,                
 J.~B\"ahr$^{36}$
% ...

\end{flushleft}
\begin{flushleft}
\it
% ...
$ ^{11}$ DESY, Hamburg, Germany$^a$ \\
% ...
$ ^{13}$ II. Institut f\"ur Experimentalphysik, Universit\"at Hamburg,
                                     Hamburg, Germany$^ a$  \\
% ...
$ ^{23}$ Physics Department, University of Manchester,
                                          Manchester, UK$^ b$\\
% ...
$ ^{25}$ Institute for Theoretical and Experimental Physics,
                                                 Moscow, Russia \\
 $ ^{26}$ Lebedev Physical Institute, Moscow, Russia$^ f$ \\
% ...
$ ^{29}$ LPNHE, Ecole Polytechnique, IN2P3-CNRS,
                             Palaiseau, France \\
 $ ^{30}$ LPNHE, Universit\'{e}s Paris VI and VII, IN2P3-CNRS,
                              Paris, France \\
% ...
$ ^{35}$ Fachbereich Physik, Bergische Universit\"at Gesamthochschule
               Wuppertal, Wuppertal, Germany$^ a$ \\
 $ ^{36}$ DESY, Institut f\"ur Hochenergiephysik,
                              Zeuthen, Germany$^ a$\\
% ...
$ ^a$ Supported by the Bundesministerium f\"ur Bildung, Wissenschaft,
        Forschung und Technologie, FRG,
        under contract numbers 6AC17P, 6AC47P, 6DO57I, 6HH17P, 6HH27I,
        6HD17I, 6HD27I, 6KI17P, 6MP17I, and 6WT87P \\
 $ ^b$ Supported by the UK Particle Physics and Astronomy Research
       Council, and formerly by the UK Science and Engineering Research
       Council \\
% ...
$ ^f$ Supported by the Deutsche Forschungsgemeinschaft \\
% ...

\end{flushleft}
\setlength{\parindent}{6mm}

\newpage
\normalsize
%

\section{Introduction}
\label{sec-introduction}

Intensity interferometry was first used in radio astronomy by
Hanbury-Brown and Twiss (HBT-effect~\cite{HBT}) to measure the spatial
extent of stars in 1953. In particle physics the HBT-effect was
rediscovered in 1959 by G. and S. Goldhaber, Lee and Pais in
antiproton-proton annihilation~\cite{GGLP1}. The GGLP effect, as it
was called then, in its original form, described the tendency of
pairs of like-sign pions to have smaller opening angles than pairs
of unlike-sign pions. This effect was shown to
arise from the symmetrisation of the wave functions of multi-boson
states, hence is also known as Bose-Einstein correlation, and now
normally refers to the enhanced probability for identical boson pairs
to have similar momenta. The shape of
the enhancement in relative momentum space was shown to be related to
the spatial dimensions of the pion source. Bose-Einstein correlations
(BEC) have since been studied in high-energy physics in nearly every
type of reaction, e.g.  in $e^+e^-$ annihilation, in
lepton nucleon ($\ell N$) scattering, in $\pi N$, $KN$, $NN$, and
$\overline{N}N$ reactions, as well as in relativistic heavy-ion
collisions with the aim of understanding the space-time evolution of the
boson source. Many reviews of this subject
exist~\cite{goldhaber,zajc,lorboquehay,dewolfglas}.  In recent years
it has been emphasized that Bose-Einstein correlations are but one of
the many facets of multi-particle correlation studies (see
e.g. \cite{dewolfvietri,dewolfpr} for a review). 
Independently of the details of the interpretation, the 
experimental study of BEC in a broad variety of reactions should
eventually lead to better understanding of the space-time geometry of
multihadron production~\cite{bjorken}.

\subsection{Bose-Einstein correlations}
\label{intro1}

Two-particle Bose-Einstein correlations are usually  described
in terms of normalized two-particle densities
\begin{equation}
R=\frac{P(1,2)}{P(1)P(2)}
\ ,\label{becf0}
\end{equation}
where $P(1,2)$ and $P(i)$ $(i=1,2)$ are, respectively, the two-particle and
single-particle inclusive densities. Bose-Einstein correlations show
up as an enhancement in the production rate of identical pions with
similar four-momenta, and reflect both geometrical and dynamical properties
of the particle radiating source. If the particles are assumed to be
emitted from a set of independently radiating sources distributed in space-time with a
distribution $\rho(\xi)$, and the particles are identical bosons,
imposing Bose symmetry on the production amplitude leads to a
correlation function
\begin{equation}
R(T) = R_0(1 + \lambda \mid \tilde{\rho}(T) \mid^2)\ ,
\label{becf00}
\end{equation}
where $\tilde{\rho}$ is the Fourier transform of $\rho(\xi)$; $R_0$ is a
normalisation constant and $T$ is related to the invariant mass $M$ of
a pion pair with four-momenta $p_1$ and $p_2$ by
\begin{equation}
T^2 = -(p_1 - p_2)^2 = M^2 - 4 m^2_{\pi }\ .
\label{fun_t} 
\end{equation}

The parameter $\lambda$ in Eq.~(\ref{becf00}) is usually assumed to
reflect the degree of incoherence of the source, i.e. $\lambda=1$ for complete 
incoherence and $\lambda=0$ for complete coherence. A static sphere
of emitters with Gaussian density of the form
\begin{equation}
\rho(\xi)=\rho(0) \exp(-\frac{\xi^2}{2r^2})\ ,
\end{equation}
where the parameter $r$ corresponds to the size of the production volume,
leads to a correlation function 
\begin{equation}
R(T) = R_0(1 + \lambda \exp(-r^2T^2))\ ,
\label{fun_goldh}
\end{equation}
usually referred to as Goldhaber parametrisation~\cite{goldhaber}.

However, the simple geometrical interpretation of the interference pattern,
based on an optical analogy is invalid when emitters move
relativistically with respect to each other, leading to strong
correlations between the space-time and momentum-energy coordinates of
emitted particles~\cite{kolehmainen,bowler0}.  Correlations of this
type arise due to the nature of inside-outside cascade
dynamics~\cite{bjorken2} as in colour-string
fragmentation~\cite{artmen}. In the interpretation of BEC by
Andersson and Hofmann~\cite{string1} and Bowler~\cite{bowler0,string2}
in the string model, the length scale measured by BEC is therefore not
related to the size of the total pion emitting source, but to the
space-time separation between production points for which the momentum
distributions still overlap. This distance is, in turn, related to the
string tension.  The model predicts an approximately exponential shape
of the correlation function
\begin{equation}
R(T) = R_0(1 + \lambda \exp(-rT))\ ,
\label{expofit1}
\end{equation} 
where $r$ is expected to be independent of the total interaction
energy.

In recent years, considerable experimental effort has been devoted to
the study of fluctuation phenomena in multiparticle production
processes~\cite{dewolfpr}.  These studies, complemented by theoretical
analyses of the (approximately) self-similar character of perturbative
QCD cascades, concentrated on the search for ``intermittency'': the
occurrence of large fluctuations in particle density arising from
scale-invariant dynamics.  This work has provided firm indications
that intermittency is strongly connected with Bose-Einstein
correlations~\cite{dewolfpr}.  Scale invariance implies that
multiparticle correlation functions exhibit power law behaviour over a
considerable range of the relevant relative distance measure (such as
$T^2$ or $M^2$) in phase space. As such, BEC from a static source do
not exhibit power law behaviour.
However, a power law is obtained if the size of the particle
source fluctuates event-by-event, and/or, if the source
itself is a self-similar (fractal-like) object extending over a large
volume~\cite{bialas}.  Power law behaviour of BEC has been observed in
several experiments ~\cite{delphi2,NA22b,UA1a}.  In these
studies, the ratio $R$ is parametrised using the form
\begin{equation}
R(M) = A + B\left(\frac{1}{M^2}\right)^\beta\ .
\label{powerfit}
\end{equation}

\subsection{Outline}

In this paper we analyse BEC in deep inelastic
positron proton scattering (DIS) at HERA. The present study
differs from previous studies of BEC in $\ell N$
scattering~\cite{emc,nun,e665be} in several aspects.  Since the data
were taken at a tenfold higher centre of mass energy, they also cover
a larger range in the kinematical quantities Bjorken$-x$ and momentum
transfer $Q^2$~\cite{f2}. The correlations in the data can therefore
be 
analysed separately as a function of $x$, $Q^2$, and hadronic centre of mass
energy $W$. In addition diffractive and non-diffractive events can be
distinguished.
 
At HERA about 10\% of the deep inelastic scattering events
exhibit a large region of rapidity in which there is no significant
hadronic activity~\cite{diffrach1,diffraczeus}. These events are usually
attributed to diffractive processes and are well described by Monte
Carlo simulations in which diffraction is modeled as deep inelastic
positron pomeron scattering where the pomeron has partonic
structure. To the best of our knowledge, BEC have never been studied
in diffractive processes. Although it is expected that the
fragmentation mechanism in these events is similar to that in
non-diffractive DIS, it nevertheless remains of interest to analyse
BEC separately for that class of events.

Aside from comparing diffractive and non-diffractive samples, a
specific goal of our analysis is to confront the correlation data with
the different interpretations referred to above. In addition we use the
Goldhaber parametrisation as a tool for the comparison with previous
experiments.

This paper is organized as follows. After a brief presentation of the
H1 detector, the event and track selections are described in
Sect.~\ref{sec-selection}. The Monte Carlo models used for
comparison are listed in Sect.~\ref{sec-montecarlo}. The details of
the analysis procedure and systematic uncertainties are given in
Sect.~\ref{sec-analysis}. The results and conclusions are presented in
Sect.~\ref{sec-results} and Sect.~\ref{sec-conclusions}, respectively.

\section{The H1 detector}
\label{sec-detector}

The data were taken with the H1 detector at the HERA $ep$ -- collider
in 1994, with 27.5 GeV positrons colliding head-on with 820 GeV
protons at a centre of mass energy $\sqrt{s}=300$ GeV. A 
detailed description of the H1
detector can be found elsewhere \cite{h1detector}. Here we briefly
describe the components relevant for this study.

The present analysis is based on the central tracking system,
closest to the interaction point, and covering polar angles
between 20$^\circ$ and 160$^\circ$, where $\theta$ is defined with
respect to the proton beam direction ($+z$ axis). Measurements of the
charge and momentum of charged particles are provided by two large
cylindrical drift chambers (central jet chambers, CJC1 and CJC2). The
chambers have wires strung parallel to the beam axis ($z-$direction)
with the drift cells inclined with respect to the radial direction,
yielding up to 56 space points.  The measured space point resolution
in the drift coordinate ($r-\phi$ plane) is 170 $\mu$m and can, by
comparing signals read out at both wire ends, achieve a resolution of
one percent of the wire length in $z$.  Two sets of thin cylindrical
$z-$drift chambers complement the measurement of charged track momenta
and are located at the inner radii of the jet chambers. The
$z-$chambers provide track elements with typically 300 $\mu$m
resolution in $z$ and $8^{\circ}$ in $\phi$.  This requires a drift
direction parallel to, and sense wires perpendicular to, the beam axis.
The central tracking system is complemented by a forward tracking
system and a backward proportional chamber (BPC) covering the polar
angle ranges $7^\circ\leq\theta\leq 25^\circ$ and
$155^\circ\leq\theta\leq 175^\circ$, respectively. The BPC combined
with a measurement of the interaction vertex position, is used to
measure the polar angle of the scattered positron to better than 1
mrad.

Surrounding the tracking system is a liquid argon calorimeter
consisting of electromagnetic and hadronic sections covering
$4^\circ\leq\theta\leq 153^\circ$ over the full azimuth.  Outside
these detectors a superconducting coil provides a uniform magnetic
field of 1.15 T parallel to the beam line.  To measure the energy of
the scattered positron, a backward electromagnetic calorimeter (BEMC)
covers the angular range $155.5^\circ\leq\theta_{e}\leq 174.5^\circ$.
Behind the BEMC is the time-of-flight system (ToF), consisting of two
planes of scintillators, with a time resolution of 1 ns. 
It is used to reject events caused by the interaction of the proton beam with
material before the detector. Such particles arrive at the
ToF system before particles from the nominal interaction point.  In
the forward direction, surrounding the beam-pipe, is the plug, a
copper/silicon calorimeter, covering the angular range
$0.7^\circ\leq\theta\leq 3.5^\circ$.  Further forward are the forward
muon drift chambers on either side of a toroidal magnet around the
beam-pipe. Their primary purpose is to detect low angle muons, but
they are also used to detect particles coming from secondary
interactions with collimators designed to protect the detector from
synchrotron radiation.

The most relevant performance parameters of the H1 detector
for this analysis are the invariant mass resolution for
charged particle pairs and the resolution of reconstructed
event kinematics (for the latter see ref.~\cite{f2}). The two-particle
invariant mass 
resolution can be deduced from the widths of the $K_S^0$ signal at $M=0.5$ GeV
and of the $J/\Psi$ peak at $M=3.1$ GeV to be $\sigma_M=8.5$ MeV~\cite{K0S} and
$\sigma_M=76$ MeV~\cite{JPsi}, respectively. This agrees with Monte
Carlo studies which also predict $\sigma_T=3.5$ MeV at $T$=0.1 GeV.

\section{Event and track selection}
\label{sec-selection}

\newcommand{\xpom}{x_{_{I\!\!P}}}
\newcommand{\mx}{M_{_X}}

The present study is based on a sample of low $Q^2$ DIS events.
The data sample comprises an integrated luminosity of 1.26
pb$^{-1}$.  The scattered positron can be detected in the BEMC for
$Q^2$ in the range from 1 GeV$^2$ to 
100 GeV$^2$.  The event selection used in this analysis follows
closely that described in our structure function analysis~\cite{f2}.
In brief, the trigger used to collect low $Q^2$ data required a
localised energy deposit in the BEMC of at least 4 GeV and no veto
from the ToF system. This has been shown to be more than 99\%
efficient for the selection of DIS events in which the scattered positron
entering the BEMC has an energy above 10 GeV.

\subsection{Event selection}
\label{ssec-dis}

All events are required to have a vertex, reconstructed from tracks in
the central tracker, located within $\pm$ 30 cm of the mean vertex
$z$--position. This cut is required to reject beam--induced background
and to permit a reliable determination of the kinematic variables. To
exclude events with large QED radiative effects and to ensure
substantial hadronic energy flow in the detector we require the invariant mass
squared, $W^2$, of the hadronic system determined both with the
Jacquet-Blondel method~\cite{jacquet-blondel} and from the scattered
positron to be larger than 4400 GeV$^2$, as well as the
scattered positron to have an energy $E_{e}\geq 12$ GeV and
a polar angle $157^\circ\leq\theta_{e}\leq 172.5^\circ$.
The energy constraint ensures that the remaining photoproduction
background is less than 1\%. Finally, at least two selected tracks are
required in each event irrespective of charge. This selection accepts
events for $Q^2$ in the range $6\leq Q^2\leq 100$ GeV$^2$ and for $x$
in the range $10^{-4}\leq x \leq 10^{-2}$. Both $x$ and $Q^2$ are 
calculated from the positron nergy and angle.

The data are divided into diffractive and non-diffractive
subsamples. Non-diffractive events are selected by requiring that the
energy deposited in the liquid argon 
calorimeter within the polar angular range $4.4^\circ\leq\theta\leq
15^\circ$ be greater than 0.5 GeV. Diffractive events are selected by
demanding that the energy deposited in the plug 
is less than 3 GeV, that there are fewer than three pairs of hits in
the forward muon drift chambers, and that the most forward significant
energy deposit (above 400 MeV) detected in the liquid argon
calorimeter has a 
pseudo-rapidity of 3.0 or less. A further cut is made using the variable 
$\xpom$, defined as
\begin{equation}
\xpom = \frac{Q^2 + \mx^2}{Q^2 + W^2}\ ,
\end{equation}
where $\mx^2$ is the mass of the hadronic final state separated from
the proton remnant system by the largest rapidity gap~\cite{diffrach1}. In order to
restrict the data to the kinematic region where pomeron exchange
dominates and to ensure good acceptance of the dissociating
pomeron-virtual boson system it is required that $\xpom \leq$
0.05. The quantity $\xpom$ can be interpreted as the fraction of the
proton momentum carried by the pomeron.

After all cuts, the non--diffractive sample contains $48\,000$ events and
the diffractive sample $2\,500$ events.  The same selections are applied
to the various Monte Carlo generated event samples used in the
analysis.  Further details are given in
references~\cite{dollfus,rizvi}.

\subsection{Track selection}
\label{sec-tsel}
 
Charged tracks are required to be linked to the primary event vertex
and to show at least 10 hits within the jet chambers. Requiring a
transverse momentum $p_t> 0.15$ GeV rejects tracks curving strongly
within the CJC. A polar angle $\theta$ satisfying $22^\circ < \theta <
150^\circ$ restricts the analysis to particles which can be measured
in both jet drift chambers in order to improve the quality of the
tracks.

BEC studies are sensitive to double counting effects, i.e. the
splitting of a single track into two pieces, which mimic two
like-charge tracks with a very low value of $T$. In this experiment we
find 
that tracks are split mainly between the inner and outer jet
chamber. Thus only tracks which start within the inner CJC are
considered. Monte Carlo studies have shown 
that this selection has a 95\% track finding efficiency, and the
remaining rate of 
split tracks is 0.3 \%. This was verified by a visual scan
of data. In addition no spurious
tracks due to noise hits or reconstructed mirror hits of real tracks
were found. To remove sensitivity to the remaining rate of split
tracks all fits to the data are performed in the region $T>0.018$~GeV.
To further assess the quality of the simulation program, a
comparison of track parameters between data and reconstructed Monte
Carlo was made. Pairs with low invariant masses were selected both in
data and in Monte Carlo and technical quantities such as radial track length,
which is a good measure of track quality, were compared. Good
agreement was found down to the lowest invariant mass included in this analysis.

All charged particles are assumed to be pions in this analysis.  
The effect of this assumption is discussed in Sect.~\ref{sec-syserr}.
The
influence of small differences in the reconstruction efficiency
between positive and negative tracks due to the geometry of the jet
chamber and a small torsion and consequent misalignment of the two jet
chambers~\cite{ref511} were taken into account by separately analysing
pairs of both charges. The effect on the results is described
in Sect.~\ref{sec-syserr} and found to be small.

\section{Monte Carlo models}
\label{sec-montecarlo}

Non-diffractive neutral current DIS events are modelled using the
Monte Carlo generator LEPTO \cite{MEPSLEPTO,lepto64}
which uses exact QCD matrix elements 
to first order to describe boson-gluon fusion, and QCD-Compton processes.
Soft parton radiation is described by either the
colour dipole model (CDM)~\cite{CDM2} as implemented by 
ARIADNE 4.03~\cite{ariadne}, in which the partons are radiated independently 
from a chain of dipoles, or the parton shower
approach~\cite{ps}, where soft parton radiation 
is generated in the ``leading log'' approximation. The
combination of matrix elements with matched parton showers is referred
to as the MEPS model~\cite{lepto64}, whilst combining matrix elements 
with CDM is referred to as MEAR and uses LEPTO 6.1~\cite{MEPSLEPTO}.

To study the effects of QED radiation the generator
DJANGO 6.0~\cite{DJANGO} is used which is an interface between
HERACLES 4.4~\cite{HERACLES}  and LEPTO. It allows the inclusion of
$\mathcal{O}(\alpha$) processes including real photon emission from the
positron as well as virtual electro-weak corrections.

The Monte Carlo generator RAPGAP 2.1~\cite{rapgap} is used to model diffractive 
DIS. The model uses parton density functions taken
from leading order QCD fits to the measured diffractive structure
function $F_2^{D(3)}$~\cite{pomeron}. As with LEPTO, exact matrix
elements are used to calculate first order QCD radiation, with
additional radiation incorporated using either CDM, or parton
showers. In addition RAPGAP includes virtual QED effects to first
order and real photon emission. The model employing RAPGAP with CDM is
referred to as RAPA, and RAPGAP with parton showers is referred to as
RAPP (Fig.~\ref{r-diff}). 

All Monte Carlo generators used in this analysis employ the Lund string
hadronisation model as implemented in JETSET 7.4~\cite{jetset}.  A
method for the simulation of Bose-Einstein correlations is 
available as an option in the JETSET Monte Carlo~\cite{luboei}. The
algorithm does not rigorously model BEC by symmetrising production
amplitudes, but treats it as a final state interaction by reshuffling
hadron momenta. As an option, the model assumes a Gaussian shape of the
Bose-Einstein enhancement\footnote{The parameters $MSTJ(51)$ and
$MSTJ(52)$ in the generator code were changed from their default
values (see \cite{jetset,luboei}) to $2$ and $7$, respectively. This choice
leads to the inclusion of BEC for multi-pion and multi-kaon states
with a Gaussian 
parametrisation, for which we have chosen $PARJ(92) = \lambda =1$ and
$PARJ(93) = (\hbar c/r)= 0.38$ GeV, for $r$=0.53 fm.} according to
the Goldhaber prescription of Eq.~(\ref{fun_goldh}). This was
used to generate a sample of non-diffractive DIS events referred to as
MEAR(BEC). Unless stated otherwise, all Monte Carlo models refer to those 
without BEC simulated.

The H1 detector simulation program, H1SIM, is based on the CERN
package GEANT~\cite{geant} and was used to simulate the detector
response in detail and to correct the data for geometrical acceptance,
kinematical cuts, resolution and particle interactions with the
detector material.  The simulated events were processed
through the same reconstruction chain as real data.

\section{Analysis procedure}
\label{sec-analysis}

\subsection{Construction of the correlation function}
\label{sec-correl}

In order to measure the correlation function of Eq.~(\ref{becf0})
we normalize the two-particle like-sign inclusive density denoted by
$\rho_2^{l}(T)$ to a reference sample $\rho^{ref} (T)$ which ideally
contains no Bose-Einstein correlations and form the ratio
\begin{equation}
R(T)=\frac{\rho_2^{l}(T)} {\rho^{ref} (T)}\ ,
\hspace*{5mm}{\rm with}\hspace*{5mm}
\rho_2^l(T)\equiv\frac{1}{N}\frac{dn^{\pm\pm}(T)}{dT}
\hspace*{5mm}{\rm and}\hspace*{5mm}
\int\rho_2^l(T)\,dT\equiv \langle n^l \rangle\ . 
\label{formula2}
\end{equation}
Here $N$ is the total number of events, $n^{\pm\pm}(T)$ the number of 
$(++)$ and $(--)$ pairs in the sample, and $<\!n^l\!>$ is the mean
number of like-sign pairs per event. 

The choice of the reference sample is not trivial and has been a
source of bias and systematic errors in all BEC measurements to date.
Ideally it should satisfy the following conditions: absence of BEC,
presence of correlations due to the topology and the global properties
of the events, and absence of dynamical correlations due to resonances
not present in like-sign pairs.  In this analysis we use either the
two-particle unlike-sign inclusive distribution
\begin{equation} \rho^{ref}(T) = \rho_2^{u}(T)
\equiv\frac{1}{N}\frac{dn^{\pm\mp}(T)}{dT} 
\hspace*{5mm}{\rm with}\hspace*{5mm}
\int\rho_2^u(T)\,dT\equiv \langle n^u \rangle\ , 
\end{equation}
where $n^{\pm\mp}(T)$ is the number of $(+-)$ pairs in the sample, and 
$\langle n^u \rangle$ is the mean number of unlike-sign pairs per event, or we
create uncorrelated pairs by mixing tracks from different events,
denoted by
\begin{equation}\rho^{ref}(T) = \rho_1\otimes\rho_1(T)\ ,
\hspace*{5mm}{\rm normalized\; as}\hspace*{5mm}
\int\rho_1\otimes\rho_1(T)\,dT = \langle n^l \rangle\ .
\end{equation}
A mixed event is constructed by taking all combinations of each track
in one event with all tracks in another event.
In order to reduce statistical errors each event is mixed with 20
others.

We then define three ratios of densities as follows:
\begin{equation}
R^{lm} = \frac{\rho_2^{l}}{\rho_1\otimes\rho_1}\ ;
\hspace{1.0cm}
R^{um} = \frac{\rho_2^{u}}{\rho_1\otimes\rho_1}\ ;
\hspace{1.0cm}
R^{lu} = \frac{\rho_2^{l}}{\rho_2^{u}}\ .
\end{equation}
The same techniques were commonly used in the past in many experiments 
\cite{delphi2,NA22b,emc,nun,aleph,tpc,amy,cleo,vepp}, but for each
experiment special treatment is necessary.

At HERA, events are observed in a wide range of virtual-photon proton
($\gamma^* p$) centre of mass energies, and hence the data sample
comprises quite different topologies depending on the position of an
event in the $x-Q^2$ plane. Events with large $Q^2$ on average have a
current (struck quark) system which is boosted further forward in 
the laboratory system than events at
lower $Q^2$. Event topologies as seen in the transverse ($r-\phi$)
plane have a hadronic system which is balanced against the 
 scattered lepton. Mixing tracks from events with
non-collinear hadronic systems, would yield track pairs with artificially
large momentum differences. Thus, in order to overcome this
bias, all events are rotated in the $r-\phi$ plane such that
the positron has $\phi=0$. A similar effect is expected to hold in the
$\theta-z$ plane. However, this effect can be reduced by mixing
events with similar $Q^2$ and $W$ values. Mixed events are, therefore, required
to differ in $Q^2$ by not more than 20 GeV$^2$, and in $W$ by less
than 20 GeV.  The difference in event charged multiplicity was restricted to
be less than three.

\begin{figure}[htb]
\centering
\epsfig{file=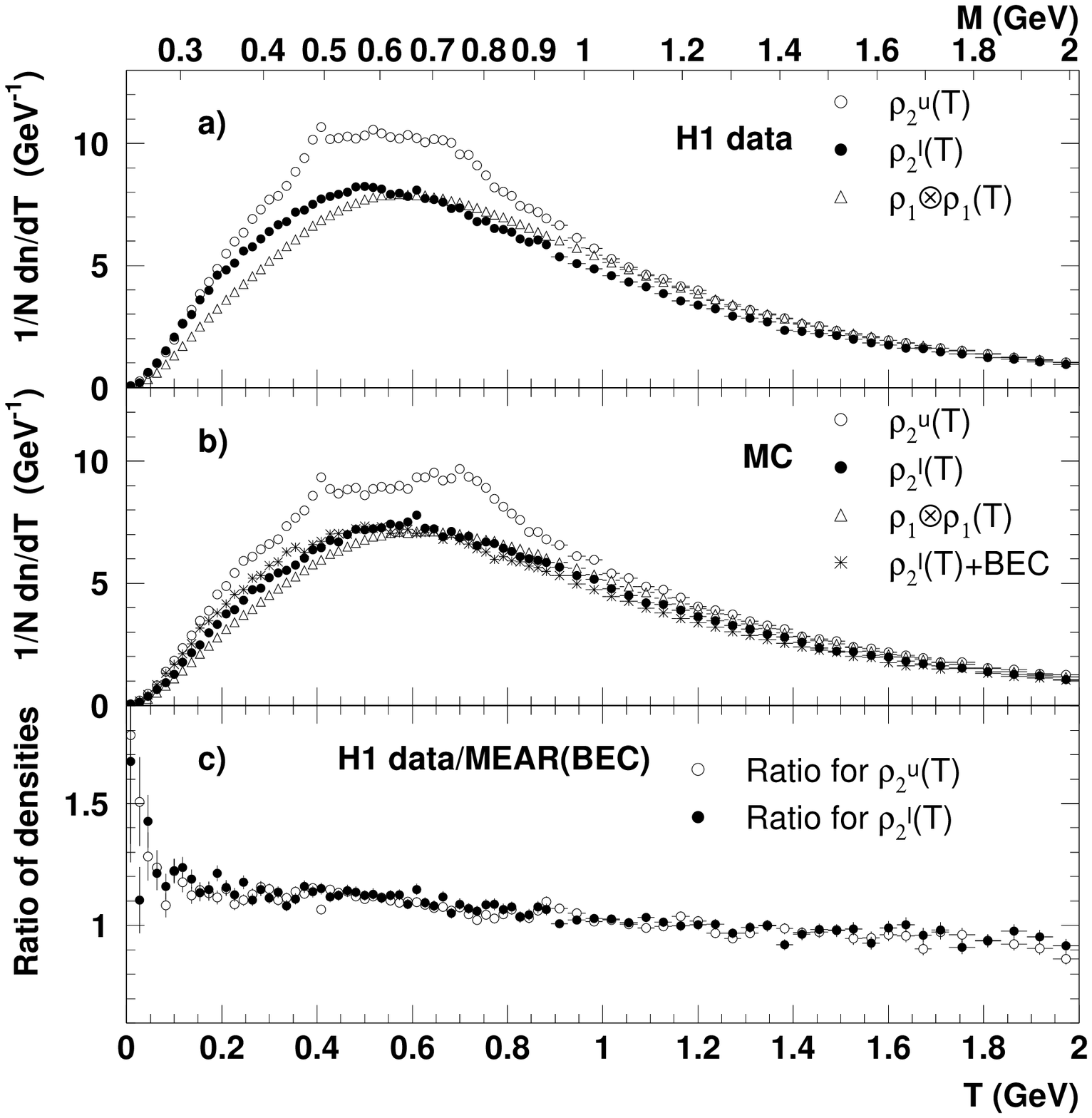,angle=0,width=\textwidth}
\caption{ Non-diffractive events - {\bf a)} Inclusive two-particle 
densities as a function
of $T$ for data: full circles denote the like-sign
pairs, $\rho_2^{l}(T)$; open circles unlike-sign pairs,
$\rho_2^{u}(T)$; open triangles the density obtained from mixing
events, $\rho_1\otimes\rho_1(T)$.  {\bf b)} Inclusive two-particle
densities for reconstructed MEAR Monte Carlo events as in (a). MEAR(BEC)
predictions for like-sign pairs are indicated as stars.  {\bf
c)} Ratios of data to reconstructed MEAR(BEC) for the like and
unlike-sign densities.}  
\label{density}
\end{figure}

The unlike-sign, like-sign and event-mixed densities constructed from our data are
shown for the non-diffractive sample in Fig.~\ref{density}a and for 
the Monte Carlo simulations in Fig.~\ref{density}b. Since the standard Monte
Carlo simulation does not include BEC, the like-sign pair distribution
constitutes, in principle, the ideal reference distribution for a
Monte Carlo with BEC included (shown as stars in
Fig.~\ref{density}b). Neither the mixed-event nor the unlike-sign
reference distribution from data is consistent with $\rho_2^l$
obtained from the Monte Carlo model without BEC. 

Mixing tracks from different events apparently shifts the distribution towards
larger $T$ compared to the invariant mass distribution of the same event, as we
can deduce by comparing $\rho_1\otimes\rho_1(T)$ with $\rho_2^{l}(T)$ ,
both shown in Fig.~\ref{density}b. We therefore conclude that despite
the precautions taken in constructing it, the mixed-pair sample 
still differs from the ideal reference distribution in the topological,
momentum and charge conservation constraints.

\begin{figure}[htb]
\centering
\epsfig{file=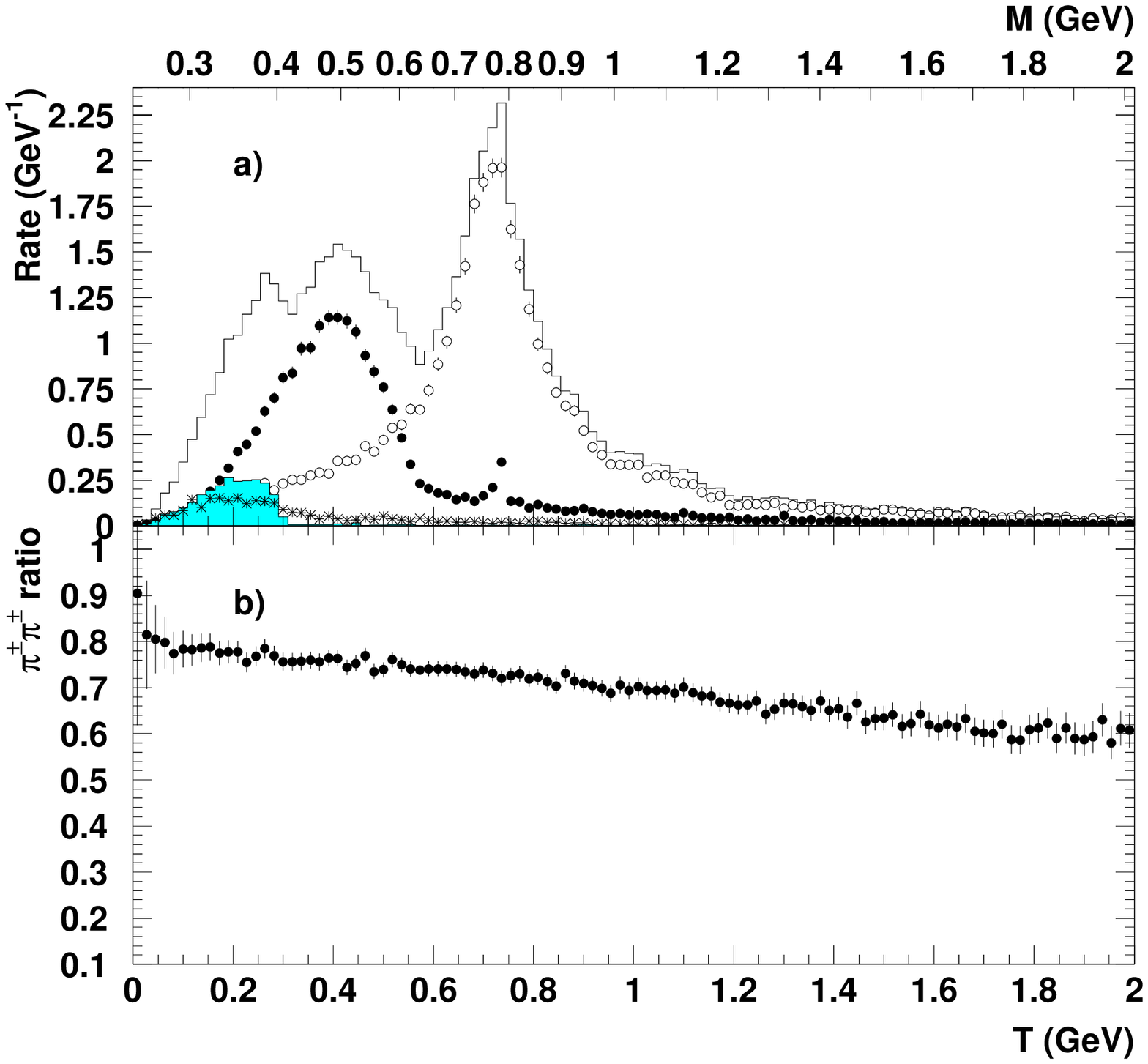,angle=0,width=\textwidth}
\caption{ {\bf a)} MEAR Monte Carlo predictions for the invariant mass
  distributions for unlike-sign pion pairs from the decays of the following
  particles:
  $\omega$ (full circles), $\rho$ (open circles) and $\eta$
  (shaded-histogram). The full histogram is the sum of all three contributions. 
  The contribution to the like-sign density from
  $\eta'$ decays is shown as stars. {\bf b)} MEAR Monte Carlo prediction for the
  ratio of densities for $\pi^\pm\pi^\pm$ pairs to that of all like-sign
  charged tracks pairs.}
\label{monteca}
\end{figure}

The unlike-pair distribution $\rho_2^u$ contains dynamical correlations from
$K_S^0$ and resonance decays, most notably from the $\rho^0$, $\eta$,
$\eta'$, and $\omega$.  A sharp peak at $T \simeq 0.4$ GeV can be seen
in $\rho_2^u$ (Fig.~\ref{density}a) which arises from the $K^0_S$ and the broad maximum at $T
\simeq 0.7$ GeV stems from the $\rho^0$ decay. The contributions from
$\eta,\;\eta'$, and $\omega$ decay do not appear as sharp peaks and are
more troublesome for the analysis than those from $K_S^0$ and
$\rho^0$, because they contribute to the low mass region, where BEC
appear. To illustrate this we have used the Monte Carlo model
to display the contributions from resonance decays in Fig.~\ref{monteca}a.  
The $K_S^0$ (not shown) and
$\rho^0$ decays cluster in a sufficiently narrow mass region, and are
removed by excluding the regions $0.37-0.43$ GeV $(K_S^0)$ and 
$0.65-0.85$ GeV ($\rho^0$) in $T$ from the analysis.

The ratio of the data to the MEAR(BEC) model shown in Fig.~\ref{density}c 
decreases smoothly with $T$ for both $\rho_2^u$ and $\rho_2^l$. 
Distinct dips or peaks are not observed in either the low mass,
the $K_S^0$ or the $\rho^0$ region. A smooth $T$ 
dependence is also observed for the ratio of the density for pion
pairs to the density for all charged particle pairs.
This ratio was obtained within the MEAR Monte Carlo model
and is shown in Fig.~\ref{monteca}b. 

The shortcomings of the reference distributions constructed from the
data can be corrected, to first order, if both the correlated and
uncorrelated distributions are divided by the corresponding Monte Carlo
distributions. Thus in order to discriminate BEC from other
dynamical correlations, we consider a double ratio by
dividing the correlation function of Eq.~(\ref{formula2}) obtained from the data, by
that obtained from reconstructed Monte Carlo events which do not
contain BEC:
\begin{equation}
RR(T) = \frac{R^{data}(T)}{R^{MC}(T)}\ .
\label{formula4}
\end{equation}

This procedure should also correct for the detector acceptance,
analysis cuts and the lack of particle identification
provided that the Monte Carlo simulation can represent
the observed distributions sufficiently well, as we will show in
Sect.~\ref{sec-results1}. The double ratio is then fitted with a
modified version of Eq.~(\ref{fun_goldh}) or Eq.~(\ref{expofit1}) in
order to extract the parameters $r$ and $\lambda$ for 
the Gaussian and the exponential parametrisations:
\begin{eqnarray}
RR(T) = & R_0(1 + a T)(1 + \lambda \exp(-r^2T^2))\ ,
\label{formula5}\\
RR(T) = & R_0(1 + a T)(1 + \lambda \exp(-rT))\ .
\label{expofit2}
\end{eqnarray}
Here $R_0$ is a global normalization factor, and the additional
parameter $a$ allows for remaining long-range correlations. 

\subsection{Systematic errors and corrections}
\label{sec-syserr}

To verify the stability of our results with respect to the assumptions
made concerning the reference distributions, and to estimate the
systematic errors arising from experimental resolutions and the
treatment of the data, we have performed a number of checks described
in the following. In order to quantitatively assess the 
influence of an assumption or a particular analysis step, the
deviations of the parameter values relative to that of the reference
fit are determined. The reference fit uses Eq.~(\ref{formula5}) for the
Gaussian parametrisation, and Eq.~(\ref{expofit2}) and
Eq.~(\ref{powerfit}) for the exponential and power law fits,
respectively. The MEAR Monte Carlo is used to calculate the double
ratios, and the fit to the non-diffractive data is performed 
in the range from 0.018 to 2 GeV in
$T$ (0.018 to 1.0 GeV in $T$ for power law fits) and
to the diffractive data in the range from 0.036 to 2 GeV.
For the unlike-sign reference distribution, the regions of the $\rho^0$
resonance and the $K^0_{S}$ are excluded from the fit (see Sect.~\ref{sec-correl}).
The observed deviations from the reference fit are then considered
in the evaluation of the systematic error. 

We first discuss those checks which use Monte Carlo
events only, and where we study the influence of reference
sample choice and detector effects.

\begin{description}

\item[Reference sample:] The systematic effects associated with the
  choice of the reference sample can be studied ``in Monte Carlo only''
  experiments using MEAR(BEC) treated as ``data''. MEAR
  without BEC is used for the denominator of the double ratios. This
  can be done at both the reconstructed level, i.e. after full detector
  simulation and reconstruction, and at the generator level. In
  addition, the like-sign density from MEAR without BEC may be
  used ideally in the denominator for $R(T)$ in Eq.~(\ref{formula2}),
  with $\rho_2^l$ in the numerator obtained from the
  MEAR(BEC) Monte Carlo model. In this case the single ratio
  $\rho_2^l$(MEAR(BEC))/$\rho_2^l$(MEAR) is fitted with the standard
  Gaussian parametrisation (Eq.~(\ref{formula5})). The results are given in
  Table~\ref{bemc} and show that a fit using the ideal reference
  distribution is able to reproduce the input value for $r$ (0.53 fm)
  within the statistical errors. The results of fits to the double
  ratios $RR^{lm}$ and $RR^{lu}$ are also given in
  Table~\ref{bemc}. Inspection of these results reveals, not
  unexpectedly (see Sect.~\ref{sec-correl} above), that neither the
  use of the unlike-sign nor the event-mixed reference distribution
  can reproduce Monte Carlo input values, not even at the generator
  level. The shift towards larger radii, between $0.18$ fm for
  unlike-sign pairs and $0.08$ fm for event-mixing is significant, and
  could, in principle, be used to correct the experimental results. We
  have not applied such a correction to allow comparison with other
  experiments (Sect.~\ref{sec-results3}) which have observed this
  shift, but have not corrected for it. \\The value extracted 
  for $\lambda$ (0.38) differs from the input value 
  ($\lambda=1$), since resonance decay products and charged particles
  other then pions are not distinguished when constructing
  the correlations. It is with respect to this value, that we
  must evaluate further differences introduced by
  the choice of the reference sample. Table~\ref{bemc} shows that these
  are sizeable. For this reason, the values of $\lambda$ found by 
  different experiments, and in different reactions, differ
  significantly and it seems hazardous to draw definitive conclusions
  from such comparisons.

\item[Detector Effects:] Finite detector resolution and acceptance
  as well as 
  imperfect pattern recognition in the reconstruction step can influence
  the BEC analysis. As Table~\ref{bemc} shows the difference in the
  extracted parameters between generator and detector level
  for the radius parameter is small, though clearly
  noticeable for $\lambda$. This comparison is repeated for all sets of 
  low $Q^2$ non-diffractive data discussed
  below separately and included as a contribution to 
  the systematic error. 
\end{description}

\begin{table}[htb]
\begin{center}
\begin{tabular}{|c|c|c|c|}
\hline
Monte & \multicolumn{3}{|c|}{Reference samples}\\\cline{2-4} 
Carlo & like-sign $\rho_2^l(T)$ 
      & unlike-sign $RR^{lu}$ 
      & event-mixed $RR^{lm}$ \\ 
Level &  $r$(fm) \hspace{1.2cm}  $\lambda$ 
      &  $r$(fm) \hspace{1.2cm}  $\lambda$ 
      &  $r$(fm) \hspace{1.2cm}  $\lambda$ \\
\hline
GEN &   $0.51\,\pm\,0.02\;\;0.38\,\pm\,0.02$ & & \\
\hline
REC & & $0.67\,\pm\,0.04\;\;0.48\,\pm\,0.04$ & 
        $0.61\,\pm\,0.04\;\;0.28\,\pm\,0.02$ \\
GEN & & $0.71\,\pm\,0.04\;\;0.57\,\pm\,0.04$ & 
        $0.61\,\pm\,0.03\;\;0.35\,\pm\,0.02$ \\
\hline
\end{tabular}
\caption {Gaussian fit results (Eq.~(\ref{formula5})) using
  MEAR(BEC) Monte Carlo (see Sect.~\ref{sec-montecarlo}) treated as
  ``data''. The input parameters for MEAR(BEC) were $r=0.53$ fm, and
  $\lambda=1$. MEAR without BEC is used as the denominator for
  construction of the double and single ratios. The entries ``GEN''
  and ``REC'' refer to using the Monte Carlo before and after the
  detector simulation and reconstruction step. Results from fits to
  the single ratio are given in the column labeled $\rho_2^l$. Results
  from fits to the double ratios $RR^{lm}$ and $RR^{lu}$ are given in
  the last two columns.}
\label{bemc}
\end{center}
\end{table}

The following contributions to the systematic error are 
evaluated from the data directly.

\begin{description}
\item[Models:] To assess the model dependency of the results we
  utilize both the MEAR 
  and MEPS Monte Carlo models for the construction of the acceptance
  corrected double ratios.

\item[Background parametrisation:] As alternatives to the linear
parametrisation of long-range correlations used in our reference fits
(Eq.~(\ref{formula5}) and Eq.~(\ref{expofit2}))
we used the following forms 
\begin{eqnarray}
RR(T) = & R_0(1 + a T +\epsilon T^2)(1 + \lambda \exp(-r^2T^2))\ ,
\label{fun_quad}\\
RR(T) = & R_0(1 + a T +\epsilon T^2)(1 + \lambda \exp(-rT))\ .
\label{fun_quadexpo}
\end{eqnarray}
Assuming that the uncorrelated background is constant over the full
range in $T$, as in Eq.~(\ref{fun_goldh}) and Eq.~(\ref{expofit1}),
leads to worse values of $\chi^2$ in most cases, and hence is not
considered. We also repeated all Gaussian and
exponential fits with an upper limit in $T$ of 1.2 GeV instead of 2
GeV. For the power law fit (Eq.~(\ref{powerfit})) the upper limit
was increased to 1.2 GeV. 
 
\item[Charge sign:] Differences in the detector response to positively
  and negatively charged particles could cause systematic shifts of
  the data. Therefore we analysed positive and negative pairs
  separately to assess the influence of such effects. For the
  diffractive data we have assumed the same error as for the
  non-diffractive sample, since further splitting of an already
  statistically small sample would lead to statistical errors far
  exceeding the systematic effect we were studying.
  
\item[Data treatment:] Technical changes such as bin width, track selection
  criteria, and the influence of the resonances (including
  the $K_S^0$ and $\rho^0$ mass region for the unlike-sign
  reference sample) were investigated and found to have no influence
  outside statistical fluctuations. In addition the restriction
  $T>0.018$ GeV was raised to 0.036 GeV and led to negligible changes
  in all fit parameters.
  
\item[Event-mixing:] The
  analysis was repeated using a simpler method of event-mixing whereby the
  data sample was subdivided into four $W$ classes and each event was
  mixed with one other event within the same class. In this
  analysis a less restrictive $W^2$ selection (as
  determined via the Jacquet-Blondel method) was used, in addition to
  an alternative track selection. The two analyses are in good
  agreement with each other~\cite{dollfus}. Furthermore, the analysis
  of the diffractive data was repeated with more stringent
  restrictions on the mixing of events. Specifically, the requirements
  were tightened to $\Delta W < 10$ GeV, and $\Delta Q^2 < 10$
  GeV$^2$. This was also found to have no influence on the
  results~\cite{rizvi}. 
  
\item[Final-state interactions:] We have not applied corrections for
  electromagnetic~\cite{fsti5,japcoul} and strong~\cite{fsti1} 
  $\pi\pi$ interactions in the final state to the data, or for the purity 
  of the pion sample, as e.g. done by the DELPHI-collaboration~\cite{delphi2}.
    We calculated the Coulomb corrections and found that they affect only the 
  region $T<0.05$ GeV, changing $\lambda$ and $r$ by
  $\delta\lambda=0.03$ and $\delta r=0.02$ fm, respectively.
  A purity correction in the sense, that one tries to correct
  the observed distribution eliminating decay products from
  long-lived resonances and charged particles other than pions,
  can be made by rescaling the data using the ratio of pions, which 
  are either directly produced or decay products of short 
  lived resonances, to that of all charged particles given by the
  Monte Carlo models. Using this procedure we have confirmed the
  observations of the DELPHI-collaboration~\cite{delphi2}, that only
  changes in the radius parameter of about 1$\sigma$ (statistical) are
  observed, whilst $\lambda$ becomes consistent with one (see~\cite{dollfus,rizvi}).
    
\item[QED radiation:] QED effects were investigated using the Monte
  Carlo model DJANGO~\cite{DJANGO} and found to be negligible for the
  ratios used in the analysis~\cite{rizvi}.

\end{description}

The most important contributions to the systematic errors for diffractive
and non-diffractive data are given in Table~\ref{errlist}
and Table~\ref{errlistdiff}, and are added in quadrature for the final errors
quoted. For the exponential and power law
parametrisations (Eq.~(\ref{expofit2}) and Eq.~(\ref{powerfit}))
the corresponding systematic uncertainties are listed in
Table~\ref{errlistexp} and Table~\ref{errlistpow}. When using the preferred
event-mixed technique, the systematic errors of $r$ are
sufficiently small to allow a meaningful comparison with other experiments and
between different subsets of our data. However, since systematic
uncertainties in the determination of $\lambda$ are as large as 60\%
in some cases, and $\lambda$ is also very sensitive to pion purity, we
place no emphasis on our measurement of $\lambda$, but include 
results in the following for completeness. 

\begin{table}[htb]
\begin{center}
\begin{tabular}{|c|c|c|}
\hline
   & unlike-sign $\rho_2^{u}(T)$ &
event-mixed $\rho_1\otimes \rho_1(T)$ \\
   &  $\delta r$(fm) \hspace{1cm}  $\delta\lambda$ & 
$\delta r$(fm) \hspace{1cm}  $\delta\lambda$ \\
\hline
Detector effects 
           &   $-$0.04 \hspace{0.8cm} $-$0.09 &
             $\pm$0.00 \hspace{0.8cm} $-$0.06 \\
Background parametrisation 
           &   $-$0.01 \hspace{0.8cm} $+  $0.01 & 
               $+$0.02 \hspace{0.8cm} $\pm$0.00 \\
Models     
           & $\pm$0.02 \hspace{0.8cm} $\pm$0.19 & 
             $\pm$0.00 \hspace{0.8cm} $\pm$0.06 \\
Track charge    
           & $\pm$0.01 \hspace{0.8cm} $\pm$0.01 & 
             $\pm$0.02 \hspace{0.8cm} $\pm$0.02 \\
\hline  
    Sum    & $+$0.02 \hspace{0.8cm} $+$0.19 & 
             $+$0.03 \hspace{0.8cm} $+$0.06 \\
(quadratic)& $-$0.05 \hspace{0.8cm} $-$0.21 & 
             $-$0.02 \hspace{0.8cm} $-$0.09 \\
\hline\hline
Reference sample $\dagger$   &   $+$0.18 \hspace{0.8cm}   $+$0.19 & 
                                 $+$0.08 \hspace{0.8cm}   $-$0.03 \\
Statistical error            & $\pm$0.04 \hspace{0.8cm} $\pm$0.03 & 
                               $\pm$0.03 \hspace{0.8cm} $\pm$0.02 \\
\hline
\end{tabular}
\caption[Contributions to the systematic errors]
{Contributions to the systematic errors for the non-diffractive data sample
and the Gaussian parametrisation (Eq.~(\ref{formula5})). The last row shows
the statistical error obtained from the reference fit for comparison.
$^\dagger$ Observed shift in the Monte Carlo extracted parameters
compared to a like-sign reference distribution without BEC (see text
and Table~\ref{bemc}). 
\label{errlist}}
\vspace*{2mm}
\end{center}
\end{table}

\begin{table}[htbp]
\begin{center}
\begin{tabular}{|c|c|c|}
\hline
   & unlike-sign $\rho_2^{u}(T)$ & event-mixed $\rho_1\otimes \rho_1(T)$ \\
   &  $\delta r$(fm) \hspace{1.2cm}  $\delta \lambda$ &
      $\delta r$(fm) \hspace{1.2cm}  $\delta \lambda$ \\
\hline
Background parametrisations  
                      & $\pm$0.04 \hspace{0.8cm} $  +$0.23 & 
                        $\pm$0.01 \hspace{0.8cm} $  +$0.13 \\
Models                & $\pm$0.03 \hspace{0.8cm} $\pm$0.11 & 
                        $\pm$0.01 \hspace{0.8cm} $\pm$0.08 \\
Track charge          & $\pm$0.01 \hspace{0.8cm} $\pm$0.01 & 
                        $\pm$0.02 \hspace{0.8cm} $\pm$0.02 \\
\hline
    Sum             &   $  +$0.05 \hspace{0.8cm} $  +$0.26 & 
                        $  +$0.02 \hspace{0.8cm} $  +$0.15 \\
   (quadratic)      &   $  -$0.05 \hspace{0.8cm} $  -$0.11 & 
                        $  -$0.03 \hspace{0.8cm} $  -$0.08 \\
\hline\hline
Statistical error &     $\pm$0.13 \hspace{0.8cm} $\pm$0.13 & 
                        $\pm$0.06 \hspace{0.8cm} $\pm$0.08 \\
\hline
\end{tabular}
\caption[Contributions to the systematic errors]
{Contributions to the systematic errors for the diffractive data
sample using the Gaussian parametrisation (\ref{formula5}). The
statistical error is that obtained from the reference fit.
\label{errlistdiff}}
\vspace*{3mm}
\begin{tabular}{|c|c|c|}
\hline
   & unlike-sign $\rho_2^u(T)$ & event-mixed $\rho_1\otimes\rho_1(T)$ \\
   &  $\delta r$(fm) \hspace{1.2cm}  $\delta \lambda$ 
   &  $\delta r$(fm) \hspace{1.2cm}  $\delta \lambda$ \\
\hline
Detector effects 
            &   $+$0.02 \hspace{0.8cm}   $+$0.15 
            &   $-$0.04 \hspace{0.8cm}   $+$0.11 \\ 
Background parametrisation  
            &   $-$0.27 \hspace{0.8cm}   $+$0.54
            &   $+$0.08 \hspace{0.8cm}   $-$0.09 \\
Models      
            &   $\pm$0.04 \hspace{0.8cm} $\pm$0.38
            &   $\pm$0.04 \hspace{0.8cm} $\pm$0.12 \\ 
Track charge  
            & $\pm$0.01 \hspace{0.8cm} $\pm$0.02 
            & $\pm$0.03 \hspace{0.8cm} $\pm$0.05 \\
\hline
   Sum      & $  +$0.05 \hspace{0.8cm} $  +$0.68 
            & $  +$0.09 \hspace{0.8cm} $  +$0.17 \\ 
(quadratic) & $  -$0.27 \hspace{0.8cm} $  -$0.38 
            & $  -$0.06 \hspace{0.8cm} $  -$0.16 \\
\hline\hline
Statistical error 
            & $\pm$0.09 \hspace{0.8cm} $\pm$0.08 
            & $\pm$0.11 \hspace{0.8cm} $\pm$0.06 \\
\hline
\end{tabular}
\caption[Contributions to the systematic errors]
{Contributions to the systematic errors for the non-diffractive data sample
using an exponential parametrisation (\ref{expofit2}). The
statistical error is that obtained from the reference fit.
\label{errlistexp}}
\vspace*{3mm}
\begin{tabular}{|c|c|c|c|c|c|c|}
\hline
&\multicolumn{3}{|c|}{unlike-sign $\rho_2^u(T)$}
&\multicolumn{3}{|c|}{event-mixed $\rho_1\otimes\rho_1(T)$}\\
   & $\delta \beta$ & $\delta B$ & $\delta A$ 
   & $\delta \beta$ & $\delta B$ & $\delta A$  \\
\hline
Detector effects
            &   $-$0.01   &   $-$0.001   &     $-$0.01 
            &   $-$0.09   &   $+$0.006   &     $-$0.01 \\
Background parametrisation
            & $  +$0.29   & $  -$0.002   &   $  +$0.02
            & $  +$0.12   & $  -$0.005   &   $  -$0.01\\
Models      
            & $\pm$0.30   & $\pm$0.003   &   $\pm$0.03
            & $\pm$0.08   & $\pm$0.003   &   $\pm$0.00\\
Track charge  
            & $\pm$0.15 &   $\pm$0.003 &   $\pm$0.00
            & $\pm$0.04 &   $\pm$0.001 &   $\pm$0.00\\
\hline
    Sum     & $  +$0.44 & $  +$0.004 & $  +$0.04
            & $  +$0.15 & $  +$0.007 & $  +$0.00\\
(quadratic) & $  -$0.34 & $  -$0.005 & $  -$0.03
            & $  -$0.13 & $  -$0.006 & $  -$0.01\\
\hline\hline
Statistical error 
            & $\pm$0.20 & $\pm$0.002 & $\pm$0.01
            & $\pm$0.15 & $\pm$0.006 & $\pm$0.01\\
\hline
\end{tabular}
\caption[Contributions to the systematic errors]
{Contributions to the systematic errors for the non-diffractive data sample
using a power law parametrisation (\ref{powerfit}). The
statistical error is that obtained from the reference fit.
\label{errlistpow}}
\end{center}
\end{table}

\section{Results}
\label{sec-results}  

\subsection{Comparison of the data with Monte Carlo predictions}
\label{sec-results1}

\begin{figure}[htb]
\centering
\epsfig{file=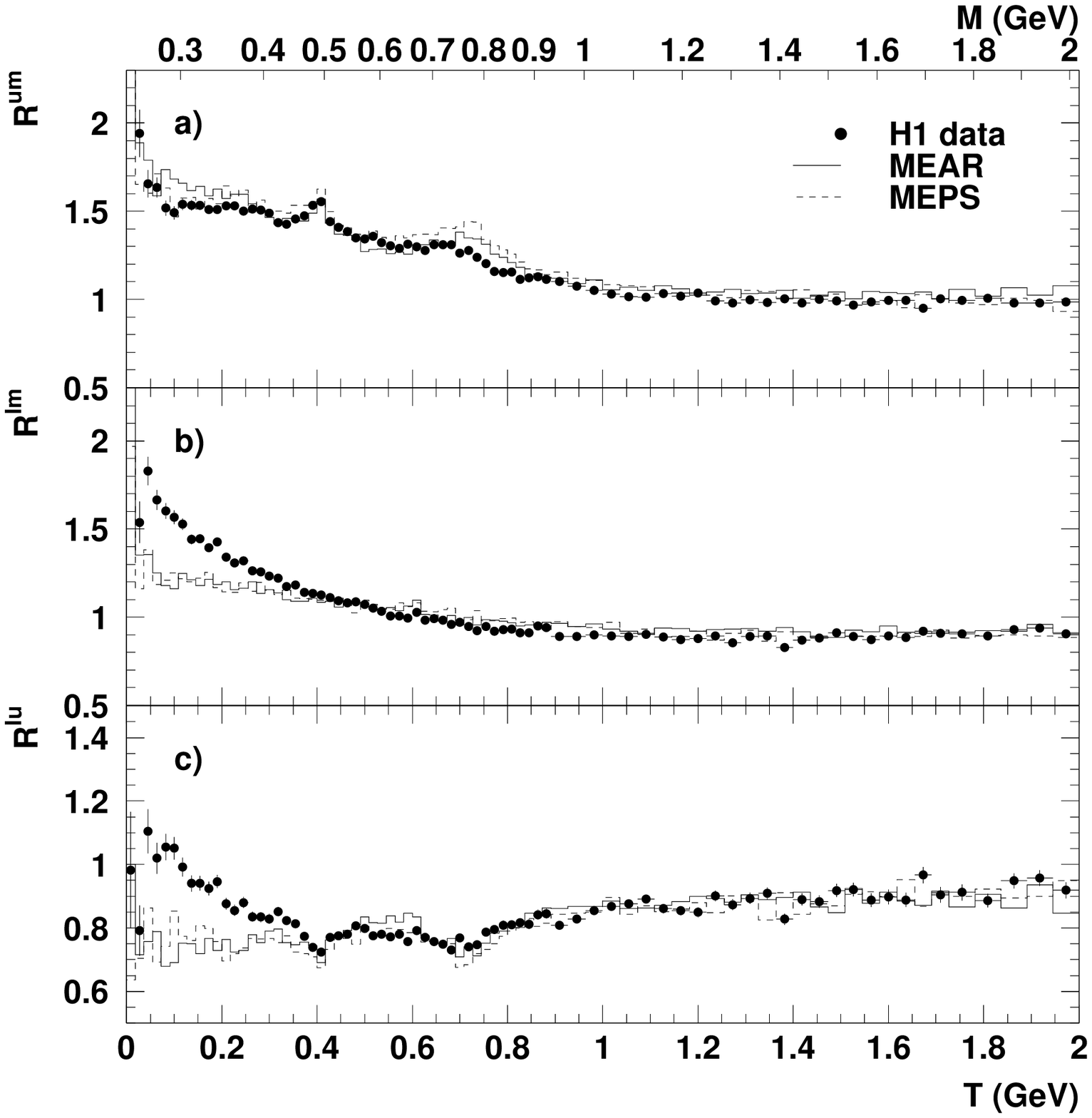,angle=0,width=\textwidth}
\caption{Comparison of the non-diffractive data (filled circles) with
the MEAR (solid line) and MEPS (dashed line) Monte Carlo simulations
for the ratios: {\bf a)} $R^{um}(T)$, {\bf b)} $R^{lm}(T)$ and {\bf
c)} $R^{lu}(T)$.}
\label{r-loq2}
\end{figure}

\begin{figure}[htb]
\centering
\epsfig{file=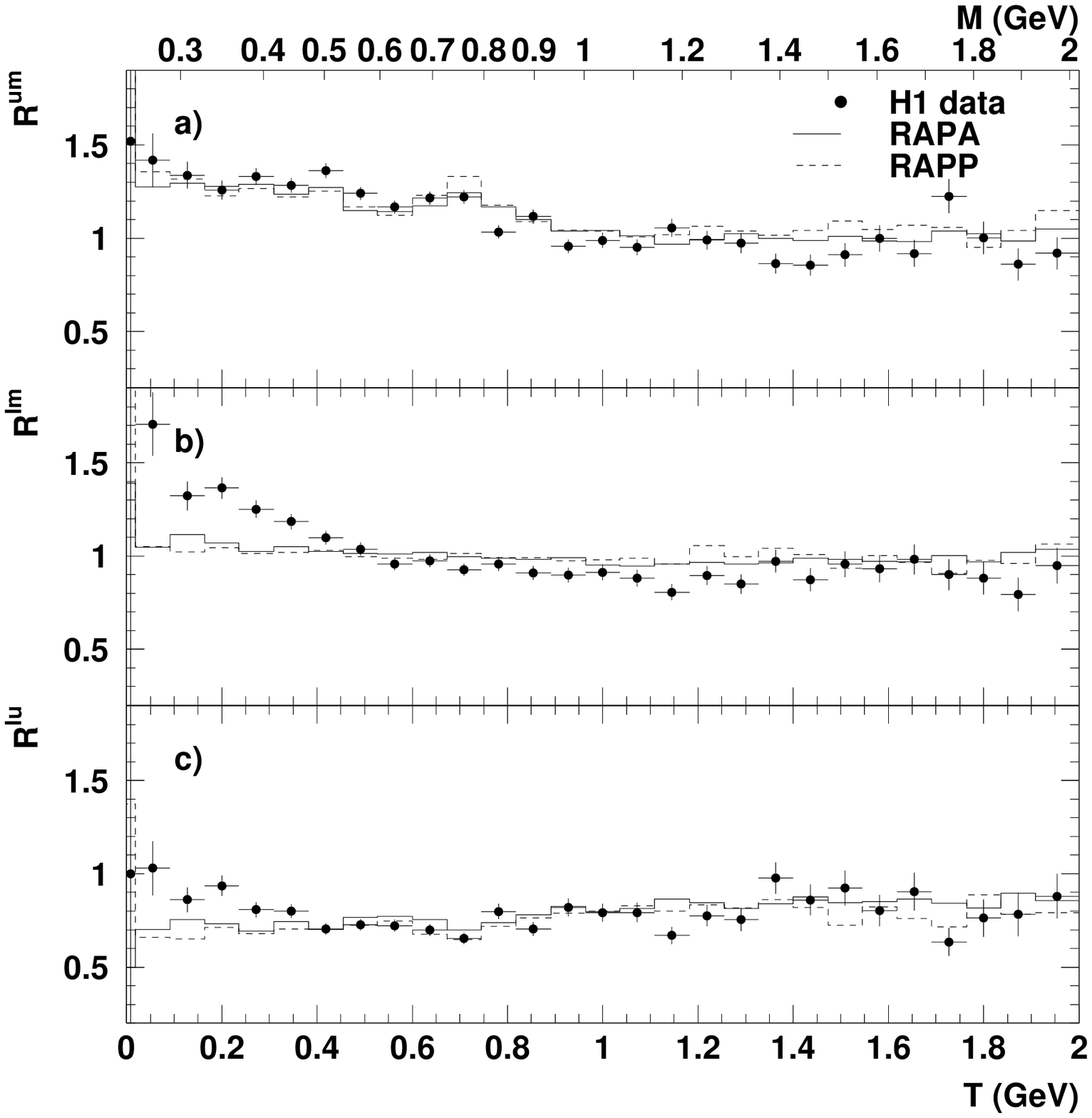,angle=0,width=\textwidth}
\caption{Comparison of the diffractive data (filled circles) with
the RAPA (solid line) and RAPP (dashed line) Monte Carlo simulations
for the ratios: {\bf a)} $R^{um}(T)$, {\bf b)} $R^{lm}(T)$ and {\bf
c)} $R^{lu}(T)$.} 
\label{r-diff}
\end{figure}

\begin{figure}[htb]
\centering
\epsfig{file=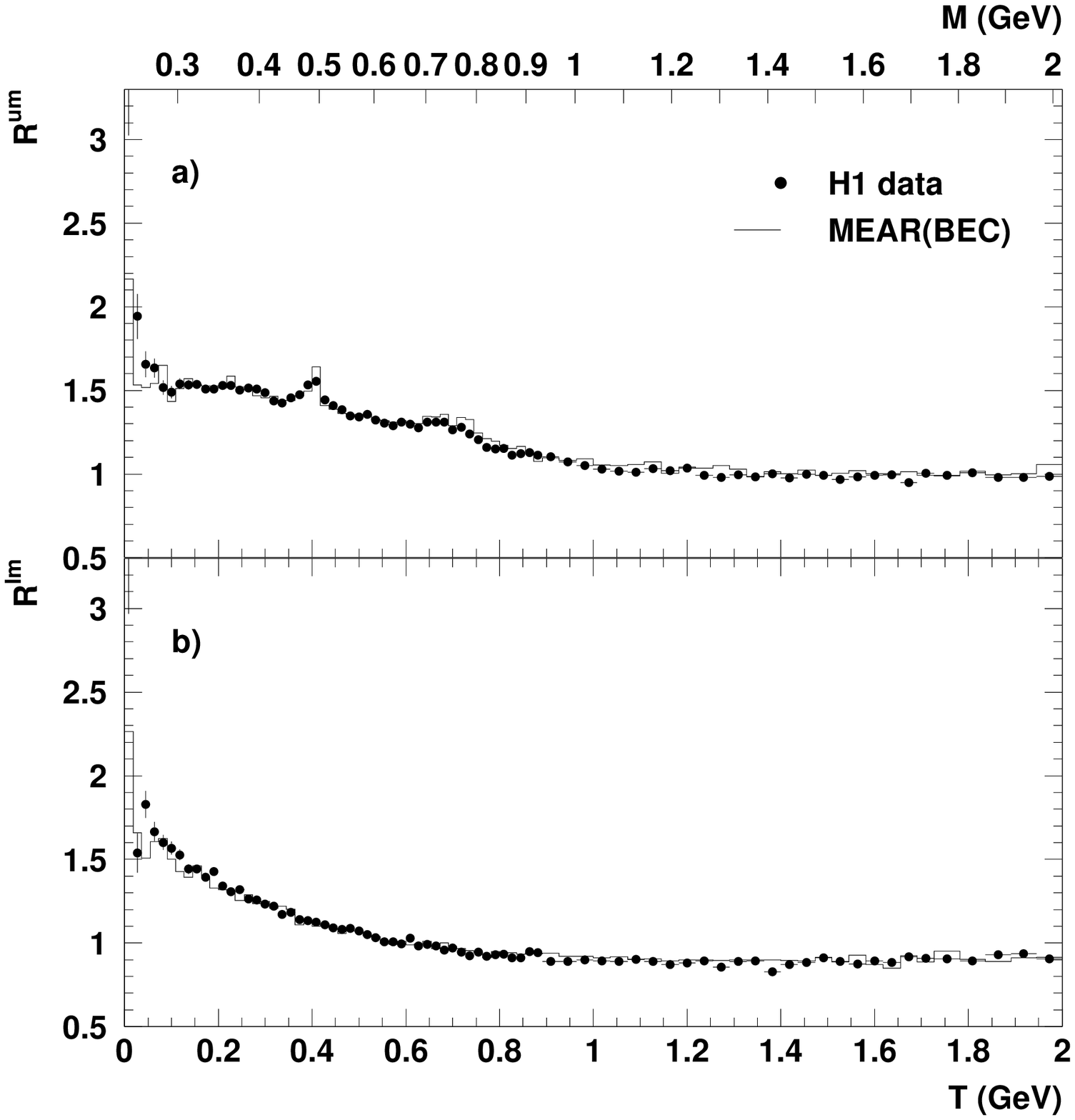,angle=0,width=\textwidth}
\caption{Comparison of the non-diffractive data (filled circles) with
the MEAR(BEC) simulation for the ratios: {\bf a)} $R^{um}(T)$
and  {\bf b)} $R^{lm}(T)$.}
\label{r-mcbe}
\end{figure}

In Figs.~\ref{r-loq2}, \ref{r-diff} and \ref{r-mcbe} we examine how
well the data are described by the Monte Carlo simulations. From the
ratio $R^{um}(T)$ shown in Fig.~\ref{r-loq2}a for data, MEAR and
MEPS Monte Carlo we can infer that the fragmentation model (JETSET)
seems to overestimate the amount of $\rho^0$ production, as previously
observed~\cite{delphi2,NA22b,aleph,NA22a}. However, as no such
discrepancy is seen in Fig.~\ref{density}c, we conclude that the
inclusion of BEC in Monte Carlo models is able to satisfactorily
describe the $\rho^0$ production region. The Monte Carlo also fails to
predict correctly the $T$ region ($0.1<T<0.2$ GeV),
where reflections of the $\eta$, $\omega$ and $\eta'$ resonances
appear (Fig.~\ref{r-loq2}a). At the smallest values of $T$ 
a sharp spike (truncated in the figures) can been seen 
in Fig.~\ref{r-loq2}a and Fig.~\ref{r-loq2}b.
For like-sign pairs this is attributed to the residual 
effects of track double counting. Similarly, photons converting into
$e^+e^-$ pairs contribute to the first bin in the unlike-sign pair
distributions. Good agreement between data and Monte Carlo model is
observed in the lowest bin (not visible in Figs.~\ref{r-loq2}a and
\ref{r-loq2}b).

In Figs.~\ref{r-loq2}b, \ref{r-loq2}c, \ref{r-diff}b and
\ref{r-diff}c the Bose-Einstein
enhancement is clearly visible for like-sign pairs for $T<0.4$ GeV in
the ratios $R^{lm}$ and $R^{lu}$. The slow rise of $R^{lm}$ towards
threshold, predicted by the models, is caused by mixing tracks of
events with a different topology as discussed in
Sect.~\ref{sec-correl} and in \cite{delphi2,aleph}.  
At larger $T$ the Monte Carlo models give a
good description of the data. The $K_S^0$ and $\rho^0$
signals are visible as distinct dips in Fig.~\ref{r-loq2}c.

In Fig.~\ref{r-mcbe}a we compare the ratio $R^{um}$ for
non-diffractive data with the MEAR(BEC) Monte Carlo simulation. Apart
from the very low $T$ region, good agreement is 
seen. Figure~\ref{r-mcbe}b shows the ratio $R^{lm}$.  Since the
radius $r$ used in the generation step lies close to the experimental 
result, this model provides a rather accurate representation of the
data except for $T<0.2$ GeV where the data systematically
exceed the prediction.  This gives evidence that the Bose-Einstein
enhancement rises somewhat faster towards threshold than expected
from a Gaussian parametrisation of the correlation function.  This was
noticed in previous experiments (see e.g.~\cite{e665be,vepp,NA22a,UA1b}).

\subsection{Fits to the measured distributions}
\label{sec-results2}

\subsubsection{BEC in diffractive and non-diffractive DIS}
\label{sec-results21}

\begin{figure}[htb]
\centering
\epsfig{file=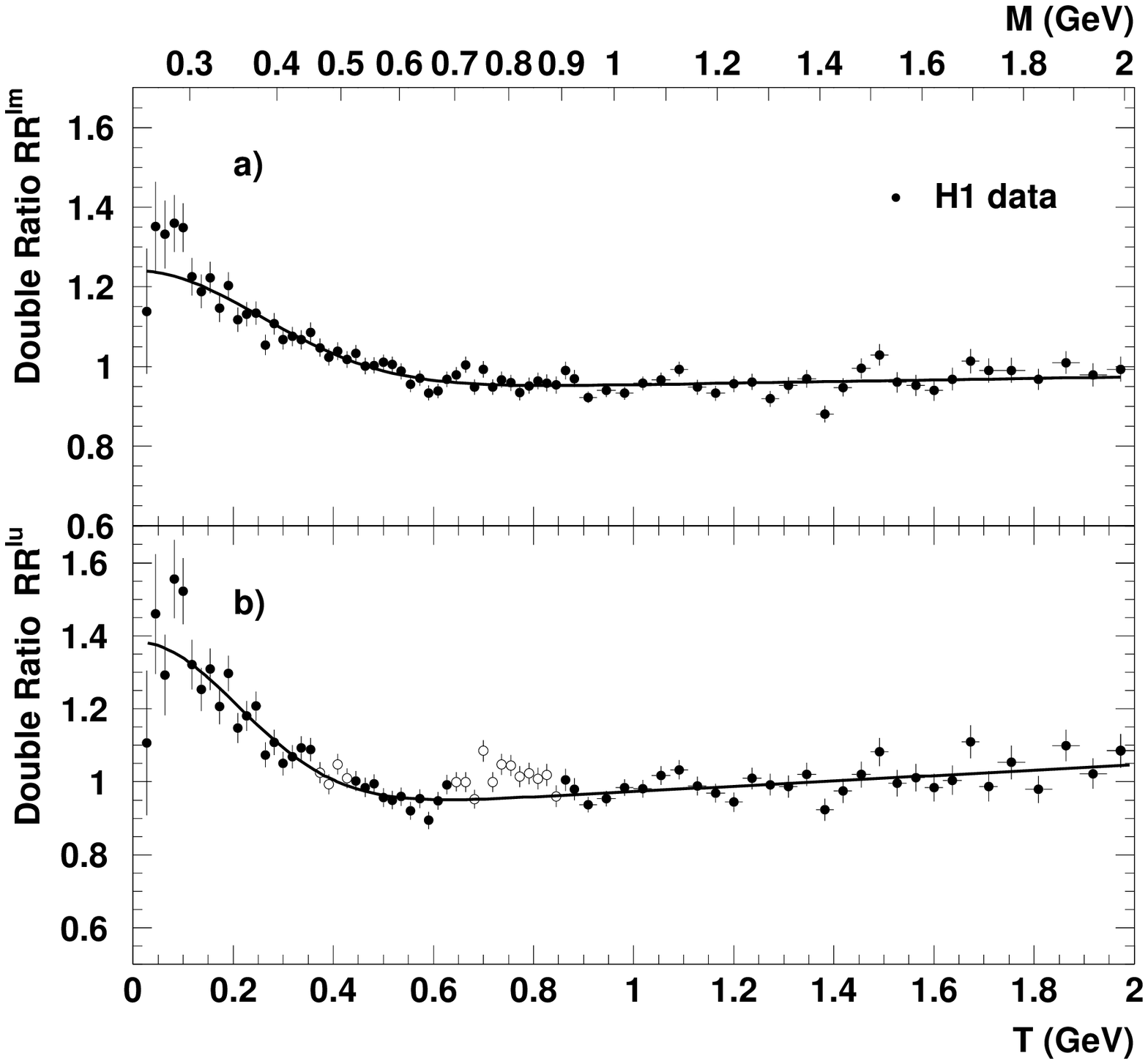,angle=0,width=\textwidth}
\caption{
{\bf a)} The non-diffractive data for the double ratio 
$RR^{lm}$.
{\bf b)} The non-diffractive data for the double ratio 
$RR^{lu}$.
The data points marked by open circles have not been included in the
fit. The solid curve results from a fit to the
Gaussian parametrisation of Eq.~(\ref{formula5}).}
\label{rr-loq2}
\end{figure}

\begin{figure}[htb]
\centering
\epsfig{file=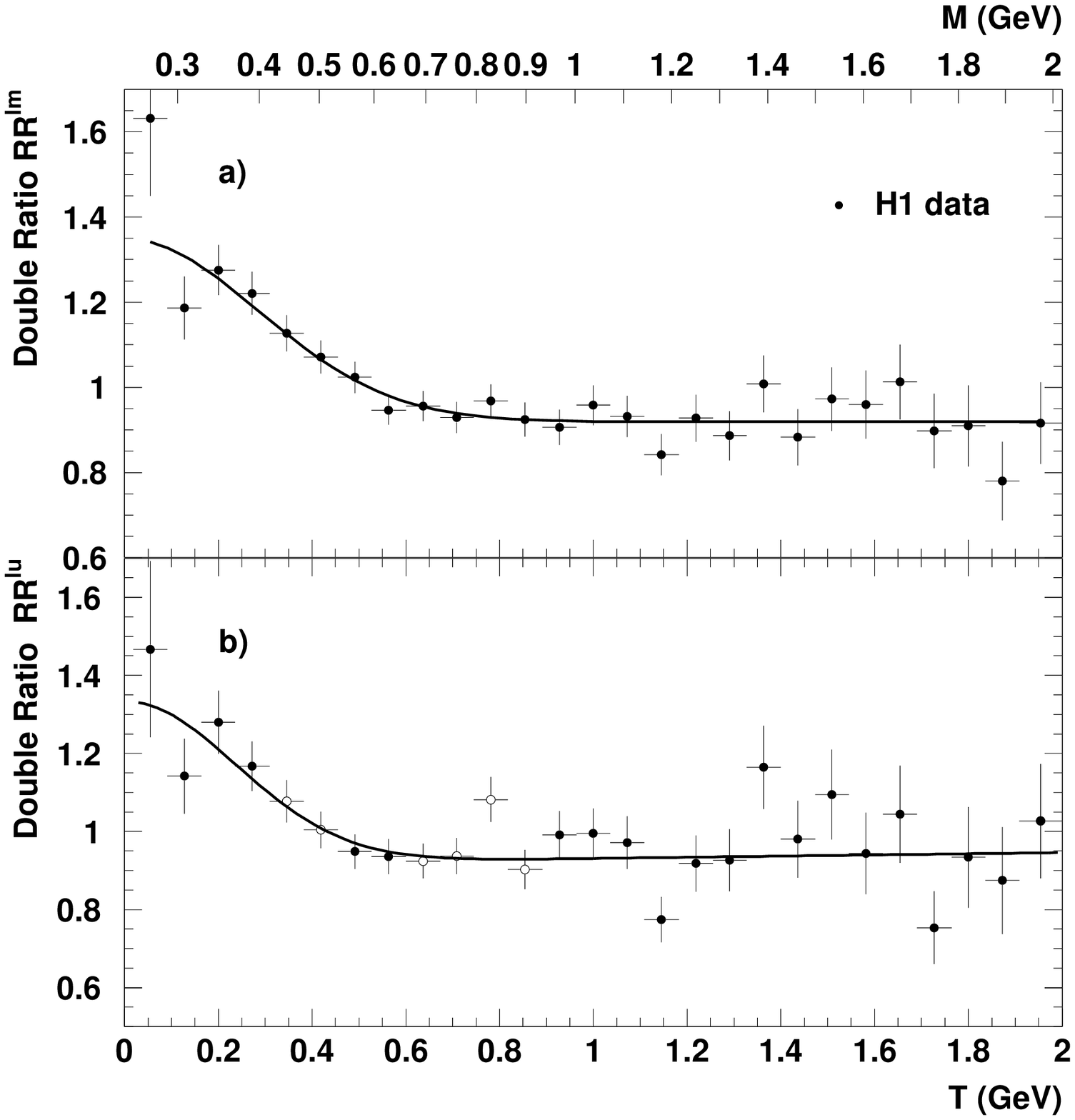,angle=0,width=\textwidth}
\caption{The diffractive data for the double ratio 
$RR^{lm}$ {\bf (a)} and $RR^{lu}$ {\bf (b)}. The data points marked by
open circles have not been included in the
fit. The solid curve results from a fit to the
Gaussian parametrisation of Eq.~(\ref{formula5}).}
\label{rr-diff}
\end{figure}

Figures~\ref{rr-loq2} and \ref{rr-diff} show the data, from which our
final results are derived and which serve as reference plots for the
discussion of the systematic errors in Sect.~\ref{sec-syserr}.  The
results from the fits to these double ratios using the Goldhaber
parametrisation (Eq.~(\ref{formula5})) are superimposed on the
data. Table~\ref{drsum} summarises the
results, which were obtained in this analysis using the 
non-diffractive and diffractive data sets, for both the event-mixed
and unlike-pair reference samples.  The parameter $a$ (typically
 between $0.01$ to $0.02$ GeV$^{-1}$) has been omitted (see
Eqs.~(\ref{formula5}) -- (\ref{expofit2})).

\begin{table}[htb]
\begin{center}
\begin{tabular}{|c|l|l|l|}\hline
Data set& \multicolumn{3}{|c|}{event-mixed $\rho_1\otimes \rho_1(T)$} \\
\cline{2-4}
& \multicolumn{1}{|c|}{$r$ (fm)} & \multicolumn{1}{|c|}{$\lambda$} & $\chi^2$/ndf \\
\hline
non-diffractive 
& 0.54 $\pm$ 0.03 $^{+0.03}_{-0.02}$  
& 0.32 $\pm$ 0.02 $^{+0.06}_{-0.09}$ & 96/72 \\
diffractive  
& 0.49 $\pm$ 0.06 $^{+0.02}_{-0.03}$
& 0.46 $\pm$ 0.08 $^{+0.15}_{-0.08}$ & 18/23\\
\hline\hline
Data set  & \multicolumn{3}{|c|}{unlike-sign $\rho_2^{u}(T)$} \\
\cline{2-4}
& \multicolumn{1}{|c|}{$r$ (fm)} & \multicolumn{1}{|c|}{$\lambda$} &  $\chi^2$/ndf\\
\hline
non-diffractive 
& 0.68 $\pm$ 0.04 $^{+0.02}_{-0.05}$
& 0.52 $\pm$ 0.03 $^{+0.19}_{-0.21}$ & 77/56 \\
diffractive  
& 0.59 $\pm$ 0.13 $^{+0.05}_{-0.05}$
& 0.46 $\pm$ 0.13 $^{+0.26}_{-0.11}$ & 26/17 \\
\hline
\end{tabular}
\caption[Summary of the two data-samples.]
{Summary of experimental results for the Gaussian fits
(Eq.~(\ref{formula5}))  
to the two data samples using both types of
reference distributions.
The first error is statistical, the second is systematic.}
\label{drsum}
\end{center}
\end{table}

As seen from Table~\ref{drsum}, the BEC parameters derived from the
event-mixed and from the unlike-sign reference samples differ
substantially. This was also observed in many earlier experiments
\cite{delphi2,emc,nun,aleph,delphi1,opal}, and subsequently ascribed
\cite{zajc,bowler} to correlations due to the production of long-lived
resonances and $\pi\pi-$interactions in the final state, as we have
discussed above.  This observation is independent of the choice of the
Monte Carlo generator used to describe the scattering process. The
difference appears already at the generator level, in the double and
single ratios as is apparent from Table~\ref{bemc}. The event-mixing
method, however, leads to values of $r$ which are in closer agreement
with the input value used in the Monte Carlo generator with BEC.

The data in Table~\ref{drsum} provide no evidence for a statistically
significant difference between the BEC parameters in
predominantly non-diffractive and diffractive DIS interactions.
The observation of similar BEC effects supports the idea that the
basic fragmentation mechanism is similar for diffractive and
non-diffractive events.

When comparing with other experiments, or searching for dependencies
on kinematic quantities such as $x$, $Q^2$, $W$, or multiplicity, we
concentrate in the following on the statistically more 
significant non-diffractive data
set.

\subsubsection{Kinematical and multiplicity dependence of BEC}
\label{sec-results22}

In order to study BEC under different conditions, the non-diffractive data
sample is divided into three classes for the three kinematic variables
separately, as listed in Table~\ref{drkinem}.  The double ratio correlation 
functions $RR(T)$ are constructed with the
mixed-event reference sample and then fitted to Eq.~(\ref{formula5}). 
The results are summarised in Table~\ref{drkinem}. The
parameters are found, within the statistical errors, to be independent
of the kinematical region considered.

\begin{table}[htb]
\begin{center}
\begin{tabular}{|c|c|c|c|}
\hline
   & $r$(fm) \hspace{1cm}  $\lambda$ 
   & $r$(fm) \hspace{1cm}  $\lambda$ 
   & $r$(fm) \hspace{1cm}  $\lambda$ \\
\hline
\hline

$x$
& 0.60$\pm$0.06  0.30$\pm$0.03 
& 0.56$\pm$0.05  0.34$\pm$0.03 
& 0.44$\pm$0.06  0.38$\pm$0.07\\
& $(0.0001\leq x<0.0006)$ & $(0.0006\leq x<0.0019)$ & $(0.0019\leq x<0.01)$ \\
\hline
\hline
$Q^2$ (GeV$^2)$ 
& 0.52$\pm$0.04  0.42$\pm$0.04
& 0.63$\pm$0.08  0.25$\pm$0.04 
& 0.47$\pm$0.04  0.41$\pm$0.05\\
& (6 $\leq Q^2 < 12$) & $(12 \leq Q^2 < 25)$ & $(25\leq Q^2 \le 100)$\\
\hline
\hline
$W$ (GeV) 
& 0.52$\pm$0.07  0.26$\pm$0.05 
& 0.48$\pm$0.03  0.42$\pm$0.04
& 0.68$\pm$0.08  0.34$\pm$0.04\\
& $(65 \leq W < 120)$  & $(120 \leq W < 180)$  & $(180 \leq W < 240)$ \\
\cline{2-4}
\hline
\end{tabular}
\caption[Double ratios with different kinematical event-classes]
{Parameters $r$ and $\lambda$ extracted using Eq.~(\ref{formula5})
for each subset of the non-diffractive data sample. Only statistical
errors are given.} 
\label{drkinem}
\end{center}
\end{table}

\begin{figure}[htb]
\centering
\epsfig{file=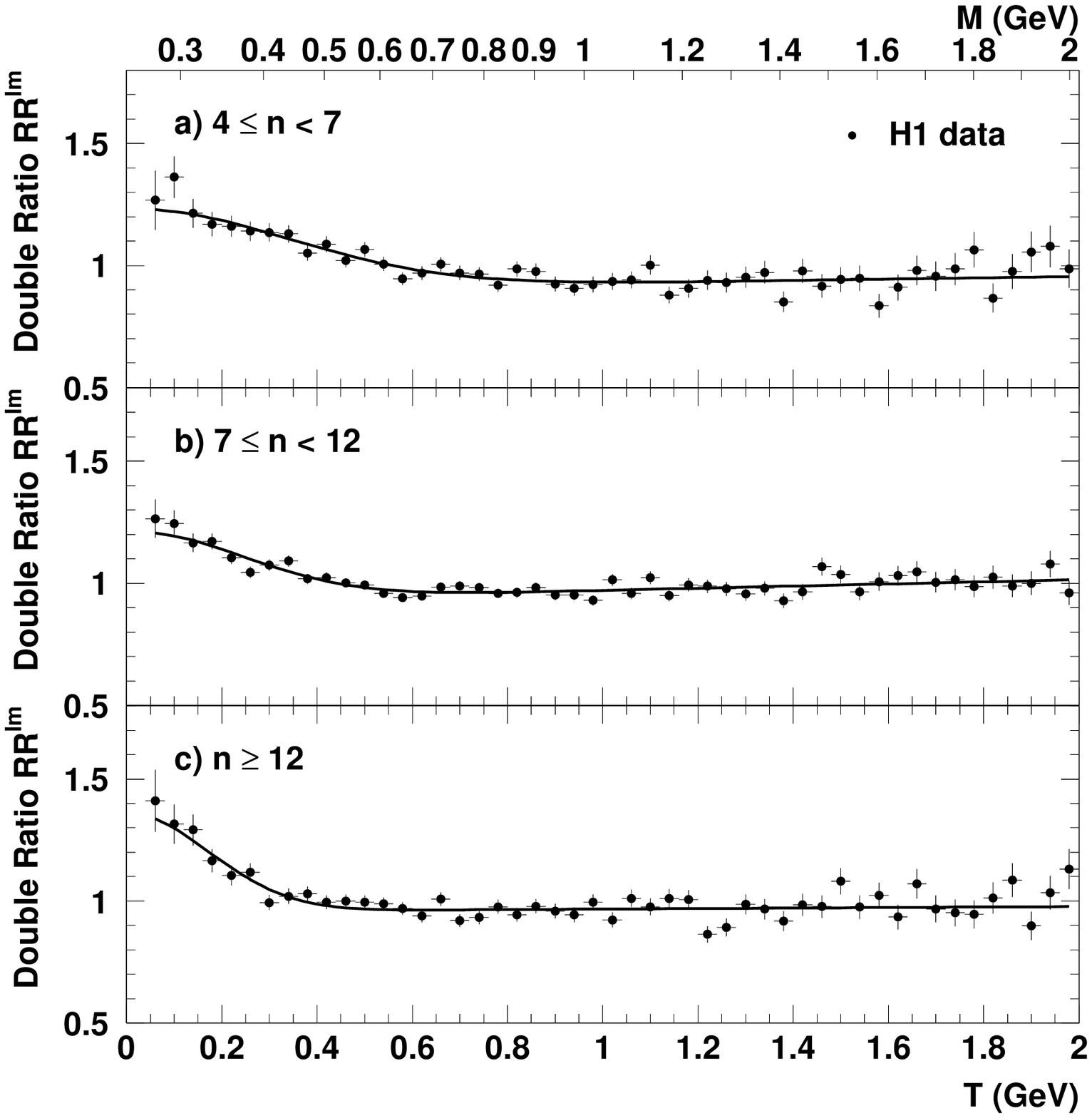,angle=0,width=\textwidth}
\caption{The double ratio $RR^{lm}$ for the non-diffractive data sample in 
intervals of {\em observed} multiplicity. The curves are fits to the
Gaussian parametrisation of Eq.~(\ref{formula5}).}
\label{kinemfit}
\end{figure}

We have also grouped the non-diffractive data into three
subsets of observed charged particle
multiplicity. Figure~\ref{kinemfit} shows the corresponding
experimental distributions and fits. The classes and the corresponding
values found for $r$ and $\lambda$ are listed in Table~\ref{drmult}.
The value of the parameter $r$ is observed to rise with increasing
multiplicity. Indications for an opposite trend are noticeable for
$x$. Previously we have observed that the multiplicity is nearly
independent of $Q^2$, but depends weakly on $W$ \cite{H1mult}. The
kinematical relation $Q^2\sim x W^2$, at small $x$, leads one then to
expect a positive correlation of $r$ with $W$ and a negative
correlation with $x$, if $r$ increases with multiplicity. The data
appear consistent with this expectation, though the effect is
not significant in view of the statistical errors.

For each bin of observed multiplicity we determine the corrected true
mean multiplicity using the iterative matrix migration method of
reference~\cite{H1mult}. The corrected mean charged particle densities
are given in 
Table~\ref{drmult}. The $\gamma^{*}p$ centre of mass pseudo-rapidity 
interval, as determined by our track selection criteria, is
$\Delta\eta^*$ = 3.2 $(1.1 < \eta^* < 4.3)$.

In order to assess the significance of the observed increase of $r$
with multiplicity, we repeat the systematic error analysis described
in Sect.~\ref{sec-syserr} for each multiplicity bin.
 The influence
of the shape of the background slope at large $T$ is such that the
systematic uncertainty (see Table~\ref{drmult}) is twice as large as the
statistical errors, and hence the rise consistent with
zero.  For comparison with recent $e^+e^-$ data~\cite{opal2} we
have repeated the analysis with the unlike-sign reference sample, and
also find a value for the slope statistically consistent with zero. 

\begin{table}[htb]
\begin{center}
\begin{tabular}{|c|c|c|c|}
\hline
Observed & Corrected & event-mixed $\rho_1\otimes\rho_1(T)$ & 
unlike-sign $\rho_2^u(T)$ \\
Multiplicity & Multiplicity 
& $r$(fm) \hspace{1cm}  $\lambda$ & $r$(fm) \hspace{1cm}  $\lambda$ \\
\hline
\hline
$4\le n<7$ & 
      4.9 $\pm$ 1.1$^\dagger$ &  
      0.42$\pm$0.05  0.37$\pm$0.05 &
      0.53$\pm$0.06  0.54$\pm$0.08\\
$7\le n<12$ & 
      8.2 $\pm$ 1.6$^\dagger$ &  
      0.58$\pm$0.05  0.31$\pm$0.03 & 
      0.77$\pm$0.07  0.54$\pm$0.06\\
$n\ge 12$ &
     13.6 $\pm$ 2.4$^\dagger$ &
      0.81$\pm$0.12  0.42$\pm$0.07 & 
      0.72$\pm$0.09  0.65$\pm$0.09\\
\hline
\hline
\multicolumn{2}{|c|}{Slope} 
& event-mixed $\rho_1\otimes\rho_1(T)$  
& unlike-sign $\rho_2^u(T)$ \\
\hline
\multicolumn{2}{|c|}{ } &&\\
\multicolumn{2}{|c|}{\Large {$\frac{1}{<r>}\frac{dr}{dn}$}}
& 0.085 $\pm$ 0.026 $^{+0.057}_{-0.048}$
& 0.045 $\pm$ 0.023 $^{+0.069}_{-0.071}$\\
\multicolumn{2}{|c|}{ } &&\\
\hline
\multicolumn{2}{|c|}{ } &&\\
\multicolumn{2}{|c|}{\Large {$\frac{1}{<\lambda>}\frac{d\lambda}{dn}$}} 
& 0.009 $\pm$ 0.018 $^{+0.048}_{-0.043}$ 
& 0.024 $\pm$ 0.026 $^{+0.024}_{-0.054}$\\
\multicolumn{2}{|c|}{ } &&\\
\hline
\end{tabular}
\caption[Double ratios with different muliplicities] {Parameters $r$
and $\lambda$ extracted using Eq.~(\ref{formula5}) for different
multiplicity subsets of the non-diffractive data sample.  The last
two rows list the result for the relative slope of the two parameters
with respect to charged multiplicity. The
first column indicates the interval in observed multiplicity, the
second column the corresponding corrected mean multiplicty.  The first
error quoted is statistical, and where given, the second is systematic. 
$^\dagger$ These numbers are not errors, but represent the spread in
true charged multiplicity in each subset.}
\label{drmult}
\end{center}
\end{table}

\subsubsection{Alternative parametrisations of BEC}
\label{sec-results23}

\begin{figure}[htb]
\centering
\epsfig{file=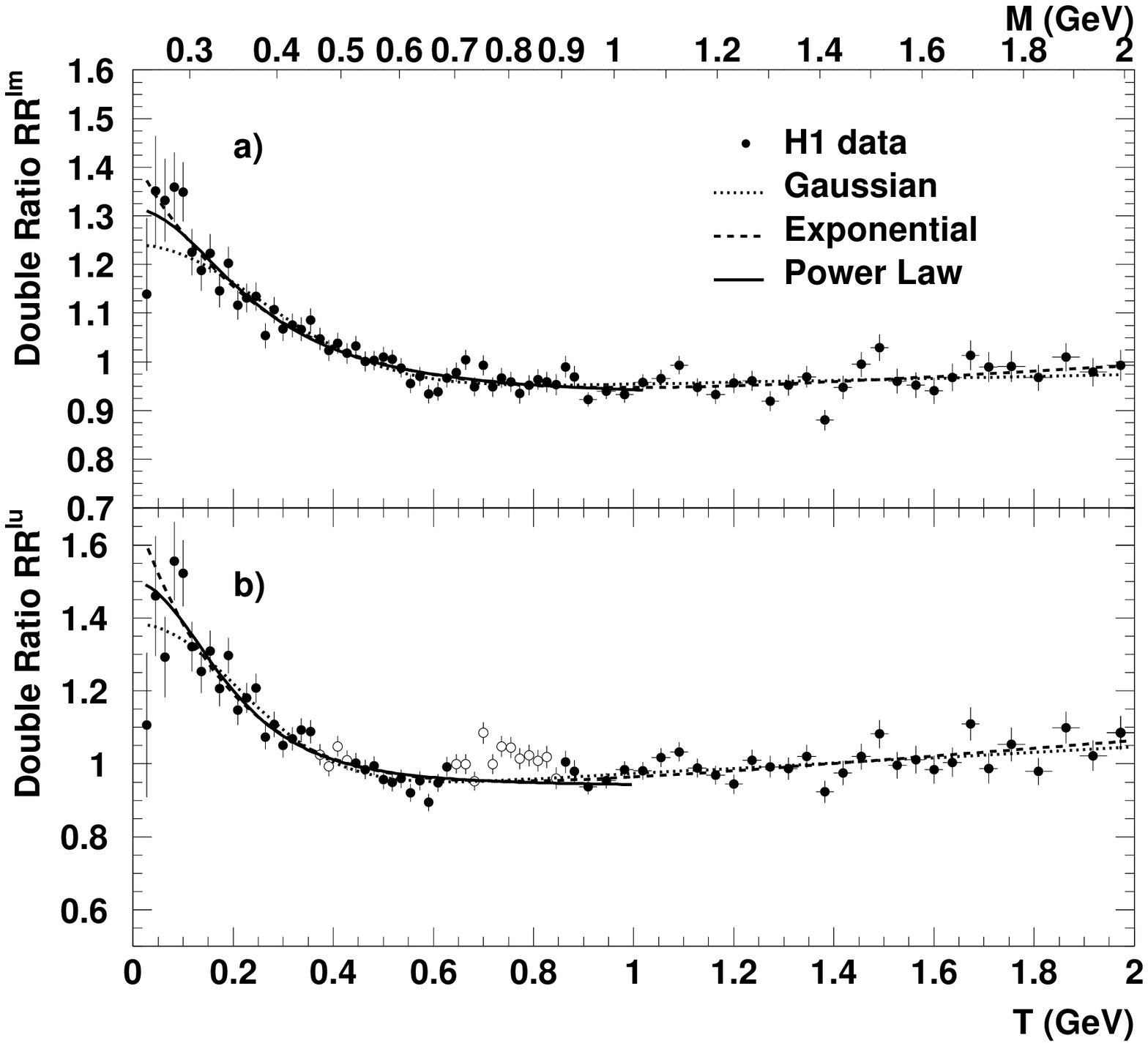,angle=0,width=\textwidth}
\caption[All fits to non-diffractive data]{The double ratios {\bf a)} $RR^{lm}(T)$, 
{\bf b)} $RR^{lu}(T)$ for the non-diffractive  data fitted to the 
Gaussian (Eq.~(\ref{formula5})), exponential 
(Eq.~(\ref{expofit2})) and power law 
(Eq.~(\ref{powerfit})) parametrisation.
\label{allfits}}
\end{figure}

Figure~\ref{allfits} shows the non-diffractive data of Fig.~\ref{rr-loq2}
together with
the results from fits using the exponential, power law, and Gaussian
parametrisations of Eq.~(\ref{expofit2}), Eq.~(\ref{powerfit}), and
Eq.~(\ref{formula5}), respectively. The parameters for the fits are
listed in Table~\ref{powerlaw}. The quality of the fit for the
exponential is slightly better than for the Gaussian parametrisation
($\chi^2/$ndf$=85/72$ compared to $96/72$) and confirms previous
observations (see e.g.~\cite{e665be}) that the Bose-Einstein
correlation function is decreasing faster with $T$ than a Gaussian.

The power law form for the two-particle correlation
function was originally suggested in the framework of a
Mueller-Regge analysis of multiparticle inclusive spectra where the
invariant mass occurs as a natural variable~\cite{berger}. In recent
studies of the intermittency effect, and its relation to BEC, power
law behaviour for like-sign particle pairs was observed in hadron-hadron
collisions~\cite{NA22b} and $e^+e^-$ annihilation at
LEP~\cite{delphi2} with values for the slope $\beta$ close to $1.5$.
From Fig.~\ref{allfits} and Table~\ref{powerlaw} it is seen that a
scale-invariant form of BEC provides a valid alternative parametrisation
of the Bose-Einstein effect. In addition we find that fitting the
power law over the range $0.018<T<2.0$ GeV, and using the fit function
\begin{equation}
RR(M) = A + \epsilon\cdot M + B\left(\frac{1}{M^2}\right)^\beta\ ,
\end{equation}
which has an additional term linear in $M$ with coefficient $\epsilon$
as a free parameter, we are able to obtain good fits
over the range $1<T<2$ GeV, too. These result in
a reduction of $\beta$ to $0.91\pm 0.18$ with $\chi^2$/ndf=86/72 for
$RR^{lm}$, and $\beta = 1.49\pm 0.19$ with $\chi^2$/ndf=80/56 for $RR^{lu}$.

\begin{table}[htb]
\begin{center}
\begin{tabular}{|c|c|c|c|c|}
\hline
Power law& $\beta$ & $B\;[$GeV$^{-2}]$  & $A$ & $\chi^2/$ndf\\
\hline
$RR^{lm}$ & 1.20  $\pm$ 0.15  $^{+0.15 }_{-0.13 }$ 
          & 0.018 $\pm$ 0.006 $^{+0.007}_{-0.006}$
          & 0.93  $\pm$ 0.01  $^{+0.00 }_{-0.01 }$ &  49/49 \\
$RR^{lu}$ & 1.82  $\pm$ 0.20  $^{+0.44 }_{-0.34 }$ 
          & 0.005 $\pm$ 0.002 $^{+0.004}_{-0.005}$
          & 0.94  $\pm$ 0.01  $^{+0.04 }_{-0.03 }$ &  52/33 \\
\hline
\hline
Exponential & $a\;[$GeV$^{-1}]$ &  $r\;[$fm$]$  
& $\lambda$ & $\chi^2/$ndf\\
\hline
$RR^{lm}$ & 0.08 $\pm$ 0.04 
          & 0.68 $\pm$ 0.11 $^{+0.09}_{-0.06}$ 
          & 0.64 $\pm$ 0.06 $^{+0.17}_{-0.16}$ & 85/72\\
$RR^{lu}$ & 0.13 $\pm$ 0.02 
          & 0.99 $\pm$ 0.09 $^{+0.05}_{-0.27}$ 
          & 1.00 $\pm$ 0.08 $^{+0.68}_{-0.38}$ & 85/56\\
\hline
\hline
Gaussian & $a\;[$GeV$^{-1}]$ &  $r\;[$fm$]$  
& $\lambda$ & $\chi^2/$ndf\\
\hline
$RR^{lm}$ & 0.02 $\pm$ 0.01 
          & 0.54 $\pm$ 0.03 $^{+0.03}_{-0.02}$  
          & 0.32 $\pm$ 0.02 $^{+0.06}_{-0.06}$ & 96/72 \\
$RR^{lu}$ & 0.08 $\pm$ 0.02 
          & 0.68 $\pm$ 0.04 $^{+0.02}_{-0.05}$  
          & 0.52 $\pm$ 0.03 $^{+0.19}_{-0.21}$ & 77/56 \\
\hline
\end{tabular}
\caption[Results from a power law fit]
{Results from the fits to the double ratios $RR^{lm}(T)$ and
$RR^{lu}(T)$ from the non-diffractive data with a 
Gaussian (Eq.~(\ref{formula5})), exponential 
(Eq.~(\ref{expofit2})) and power law 
(Eq.~(\ref{powerfit})) parametrisation. Where given, the first error
is statistical, the second systematical.\label{powerlaw}}
\end{center}
\end{table}

\subsection{Comparison with other experiments}
\label{sec-results3}

Table~\ref{drexperiment} summarises the Gaussian fit results from
other experiments for comparison with the present analysis.
For completeness also results are included, which
used correction methods different from ours, e.g. those
where purity or Coulomb corrections were applied.  These corrections
mainly influence the parameter $\lambda$, which may explain the larger
differences observed here.  The values of $r$ extracted with a mixed-event
reference sample are systematically lower than those obtained with an
unlike-sign reference, consistent with our data. However, when
allowance is made for systematic errors, which not all experiments
have quoted, but which are likely to be of similar size to ours, the measured
radii are relatively consistent with each other. The fact that the data from
$\ell N-$scattering are independent of the energy, and furthermore
agree with those from $e^+e^-$ annihilation at various centre of mass
energies (with the exception of the MARK II - data \cite{markII}),
seems to lend support to the string model interpretation, where the
source dimension is related to the string tension and is energy independent. 

The multiplicity dependence has also been studied in a few
experiments.  In $e^+e^-$ annihilation at LEP~\cite{opal2} a small
rise of $r$ with observed multiplicity has been reported by OPAL, 
while $\lambda$ was seen to decrease:
$$\frac{1}{<r>}\frac{dr}{dn}~=~0.0036~\pm~0.0006\ ,\hspace*{10mm}
\frac{1}{<\lambda>}\frac{d\lambda}{dn}~=~-0.0042~\pm~0.0008\ .$$ 
The unlike-sign reference sample was used. Our values for this reference 
sample, 
$$\frac{1}{<r>}\frac{dr}{dn}=0.045\pm 0.023\ ,\hspace*{10mm}
\frac{1}{<\lambda>}\frac{d\lambda}{dn}= 0.024 \pm 0.026\ ,$$
are consistent with these results, albeit within large errors.

It may be argued that a more appropriate comparison is to examine the slopes 
with respect to mean particle density per interval of pseudorapidity
rather then per interval of multiplicity, because the angular 
acceptance of the LEP experiments is different in the 
centre of mass system and the total multiplicity larger ($10 < n < 40$).
We have scaled the $e^+e^-$
results by the range in $\eta^*$ accessible to OPAL ($\Delta\eta^*=2$
and $\mid \eta^*\mid < 1$) and also our data by $\Delta\eta^*$ to obtain:
$$\frac{1}{<r>}\frac{dr}{d(dn/d\eta^*)}=0.0072~\pm~0.0012\;({\rm OPAL})\ ,\;\;\;
\frac{1}{<r>}\frac{dr}{d(dn/d\eta^*)}=0.144~\pm~0.074^{+0.220}_{-0.227}
\;({\rm H1})\ .$$
The two data sets are still consistent.

\begin{figure}[htb]
\centering
\epsfig{file=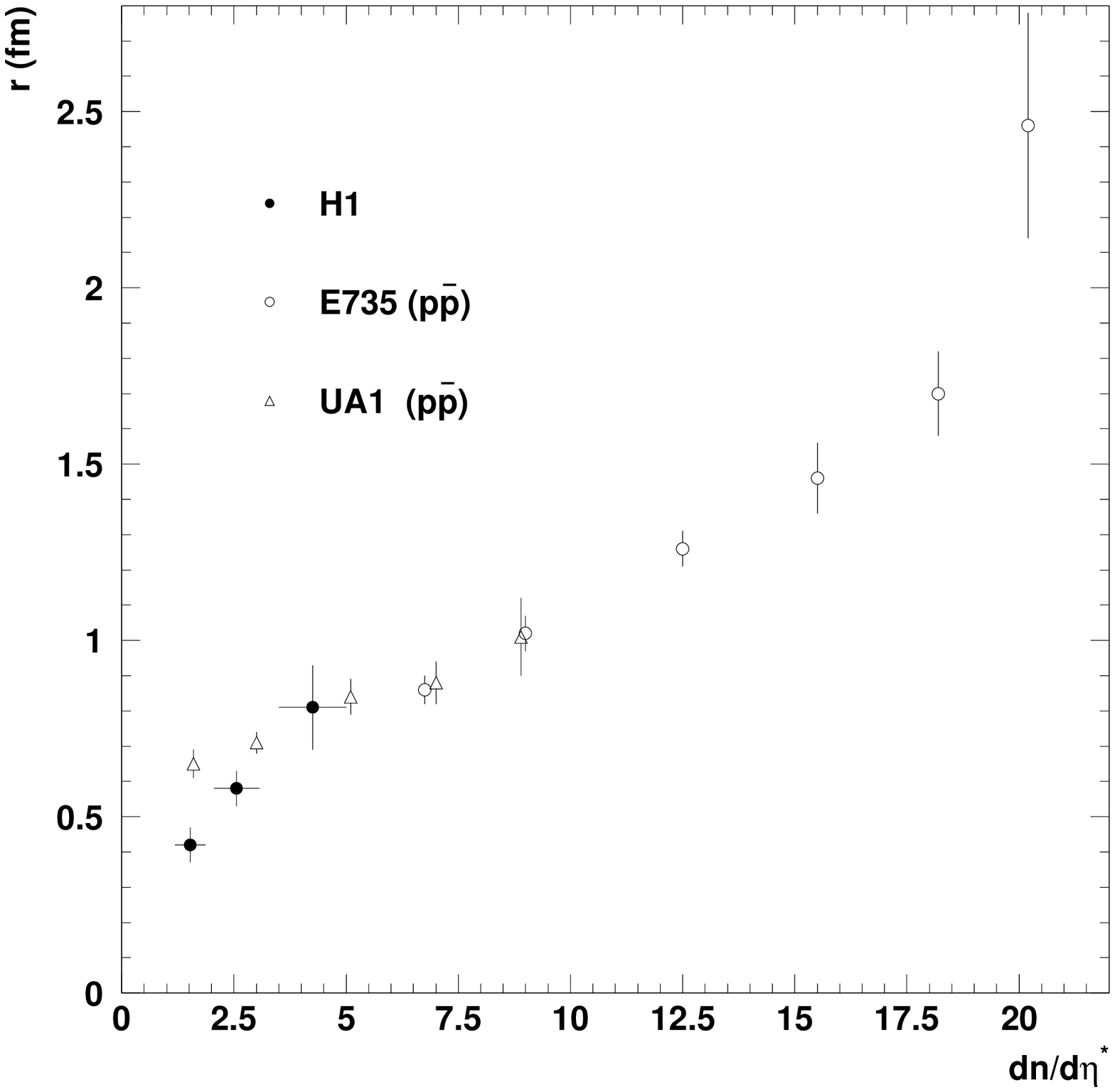,angle=0,width=\textwidth}
\caption[dn/deta]{The radius parameter $r$ versus
  charged  particle density. The full circles are from this analysis,
  the open circles (triangles) are from \pp~ collisions at a centre of
  mass energy of 1.8 TeV~\cite{E735} and 630 GeV~\cite{ua1}. Only
  statistical errors are shown. 
\label{multipli}}
\end{figure}

In proton-proton and proton-antiproton collisions, first indications
for a multiplicity dependence were seen at low centre of
mass energies at the ISR~\cite{afs}. Both the
UA1-collaboration~\cite{ua1} at centre of mass energies between 200
and 900 GeV ($\mid\eta^*\mid<3$) as well as the
E735-col\-la\-bo\-ra\-tion~\cite{E735} at 1.8 TeV ($\mid\eta^*\mid<3.25$) 
report a strong increase of radius with
particle density, $dn/d\eta^*$, in $\bar{p}p$ collisions.
Figure~\ref{multipli} shows the results on $r$, together with our data
converted into particle density, where all data were obtained using
event-mixed reference samples. For UA1 we have only plotted the
statistically most significant data set at 630 GeV. Our results match
these data at the lower end reasonably well.

\begin{table}[htbp]
\begin{center}
\small
\begin{tabular}{|l|ll|ll|}
\hline
\multicolumn{1}{|c|}{Experiment} & \multicolumn{2}{|c|}{unlike-sign $\rho_2^u(T)$} &
\multicolumn{2}{|c|}{event-mixed $\rho_1\otimes \rho_1(T)$}\\
\multicolumn{1}{|l|}{$\sqrt{s},\,W$ $[$GeV$]$}&  
$r$(fm) &  $\lambda$ & $r$(fm) & $\lambda$ \\
\hline 
\multicolumn{1}{|c|}{$e^+e^-$} &&&&\\\cline{1-1}
\multicolumn{1}{|c|}{DELPHI~\cite{delphi1}} & 0.83$\pm$0.03 & 0.31$\pm$0.02 
& 0.47$\pm$0.03 & 0.24$\pm$0.02\\ 
91 & (0.82$\pm$0.03) & (0.45$\pm$0.02) 
& (0.42$\pm$0.04) & (0.35$\pm$0.04) \\
 \multicolumn{1}{|c|}{DELPHI~\cite{delphi2}} & & & 
(0.49$\pm$0.01$\pm$0.05) & (1.06$\pm$0.05$\pm$0.16) \\
\multicolumn{1}{|c|}{ALEPH~\cite{aleph}} & 0.82$\pm$0.04 & 0.48$\pm$0.03  
& 0.52$\pm$0.02& 0.30$\pm$0.01 \\
91  & (0.82$\pm$0.04) & 
(0.62$\pm$0.04) & (0.50$\pm$0.02)& (0.40$\pm$0.02) \\
\multicolumn{1}{|c|}{OPAL~\cite{opal2}}&0.955$\pm$0.012$\pm$0.015& 0.672$\pm$0.013$\pm$0.024&& \\
91 &  & & &\\
\multicolumn{1}{|c|}{AMY~\cite{amy}}&0.73$\pm$0.05$\pm$0.20 & 0.47$\pm$0.05$\pm$0.05 &
0.58$\pm$0.06$\pm$0.02 &0.39$\pm$0.04$\pm$0.03\\58 & & & &\\
\multicolumn{1}{|c|}{TASSO~\cite{tasso}} &0.82$\pm$0.06$\pm$0.04&0.35$\pm$0.03 && \\
34  & (0.80$\pm$0.06$\pm$0.04) & (0.70$\pm$0.06$\pm$0.09) &&\\
\multicolumn{1}{|c|}{MARK II$^{\dagger}$~\cite{markII}}& 0.75$\pm$0.03$\pm$0.04 
& 0.28$\pm$0.02$\pm$0.04 & 
0.97$\pm$0.10$\pm$0.05 & 0.27$\pm$0.04$\pm$0.02\\29
& (0.84$\pm$0.06$\pm$0.05) & (0.50$\pm$0.03$\pm$0.04) & 
(1.01$\pm$0.09$\pm$0.06)&(0.45$\pm$0.03$\pm$0.04)\\
\multicolumn{1}{|c|}{TPC$^{\dagger}$~\cite{tpc}}& & & 
0.65$\pm$0.04$\pm$0.05 & 0.50$\pm$0.04 \\
29 & & & (0.65$\pm$0.04$\pm$0.05) & (0.61$\pm$0.05$\pm$0.06)\\ 
\hline
\multicolumn{1}{|c|}{$\ell N$ } &&&&\\\cline{1-1} 
\multicolumn{1}{|c|}{$\mu p$: EMC~\cite{emc}} & 0.84$\pm$0.03 & 1.08$\pm$0.10 
& 0.46$\pm$0.03 & 0.73$\pm$0.06\\$23,4\!<\!W\!\!<\!20$  &&&&\\
\multicolumn{1}{|c|}{$\mu N$: E665~\cite{e665be}} &&& 0.39$\pm$0.02 & 0.35$\pm$0.02\\
$30,8\!<\!W\!\!<\!32$ &&&&\\
\multicolumn{1}{|c|}{$\nu N$: BBCNC~\cite{nun}} 
& 0.80$\pm$0.04 & 0.61$\pm$0.04 
& 0.64$\pm$0.04 & 0.46$\pm$0.03 \\ $>\!10,2.5\!<\!W\!\!<\!40$ &&&&\\
\multicolumn{1}{|c|}{$ep$: H1} & 0.68$\pm$0.04$^{+0.02}_{-0.05}$ 
& 0.52$\pm$0.03$^{+0.19}_{-0.21}$ &
0.54$\pm$0.03$^{+0.03}_{-0.02}$ & 0.32$\pm$0.02$\pm$0.06 \\$300,W\!\geq\! 66$ &&&&\\
\hline
\end{tabular}
\normalsize
\caption[Results from other experiments] 
{Gaussian fit results for $e^+e^-$ collider experiments with $\sqrt{s}\geq 29$
GeV, and from $\ell N$ scattering experiments at high energies. 
Where two errors are given the first one is
statistical, the second systematical. All entries refer to fits to the
double ratios $RR(T)$ without Coulomb and final state
corrections. Where available, corrected results are given in brackets.
$^{\dagger}$ Values derived from the single ratios $R(T)$.}
\label{drexperiment}
\end{center}
\end{table}

\section{Conclusions}
\label{sec-conclusions}   

We have presented a study of Bose-Einstein correlations (BEC) at the highest
centre of mass energy for $ep-$collisions available so far. Even at
this high energy (300 GeV) the apparent source dimensions, measured by
the width of the BEC signal are in good agreement with data from $\ell
N$ scattering at lower energies (10 to 30 GeV) and with most data
from $e^+e^-$ reactions between $\sqrt{s}=29$ and $\sqrt{s}=91$ GeV. BEC in
diffractive DIS have been measured for the first time, and no
difference between diffractive and non-diffractive data has been
observed.  
No $Q^2$, $x$ or $W$ dependence is observed in the kinematical
ranges accessible at HERA. These observations fit well within the picture inherent
to string models, that hadronisation is a universal phenomenon and
that the measured spatial dimensions reflect mainly the constant string
tension.

For the non-diffractive data, where statistics are highest, we have
examined alternative parametrisations of the Bose-Einstein correlation
function. Both an exponential form in $T$ and a power law in
invariant mass describe the Bose-Einstein enhancement at
threshold well. The latter observation confirms, in a new energy regime,
earlier work on the search for scale-invariance in multi-hadron production.

Though the evidence for a particle density dependence of the source size,
when taken from our data alone is marginal, it seems consistent with the
trend observed in hadron-hadron collisions.


{\bf Acknowledgements:} We are grateful to the HERA machine group
whose outstanding efforts made this experiment possible. We appreciate
the immense effort of the engineers and technicians who contributed to
the construction and the maintainance of the detector.  We thank the
funding agencies for financial support. We acknowledge the support of
the DESY technical staff. We also wish to thank the DESY directorate
for the hospitality extended to the non-DESY members of the
collaboration.


\end{document}